\documentclass{aa}  

\usepackage{graphicx}
\usepackage{txfonts}
\usepackage[hidelinks]{hyperref}
\usepackage{float}
\usepackage{tikz}

\usetikzlibrary{shapes.geometric, arrows, calc}
\tikzstyle{data} = [rectangle, rounded corners, minimum width=2.0cm, minimum height=1cm,text centered, draw=black, fill=green!20]
\tikzstyle{outofscope} = [rectangle, rounded corners, minimum width=1.5cm, minimum height=1cm,text centered, draw=black]
\tikzstyle{aas2rto} = [rectangle, rounded corners, minimum width=1.5cm, minimum height=1cm,text centered, draw=black, fill=blue!20]
\tikzstyle{arrow} = [thick,-latex]

\definecolor{lime}{HTML}{A6CE39}
\DeclareRobustCommand{\orcidicon}{
	\begin{tikzpicture}
	\draw[lime, fill=lime] (0,0) 
	circle [radius=0.16] 
	node[white] {{\fontfamily{qag}\selectfont \tiny iD}};
	\draw[white, fill=white] (-0.0625,0.095) 
	circle [radius=0.007];
	\end{tikzpicture}
	\hspace{-2mm}
}

\newcommand{\orcid}[1]{\href{https://orcid.org/#1}{\textcolor[HTML]{A6CE39}{\orcidicon}}}

\newcommand{\orcidSedgewick}{\orcid{0000-0002-9158-750X}}
\newcommand{\orcidGall}{\orcid{0000-0002-8526-3963}}
\newcommand{\orcidIzzo}{\orcid{0000-0001-9695-8472}}
\newcommand{\orcidAgnello}{\orcid{0000-0001-9775-0331}}
\newcommand{\orcidAngus}{\orcid{0000-0002-4269-7999}}
\newcommand{\orcidHjorth}{\orcid{0000-0002-4571-2306}}
\newcommand{\orcidKadela}{\orcid{0009-0004-7112-0939}}

\begin{document}

  \title{AAS2RTO: Automated Alert Streams to Real-Time Observations}

  \subtitle{Preparing for rapid follow-up of transient objects in the era of LSST}

  \author{
    Aidan~Sedgewick\orcidSedgewick \inst{1}\thanks{\email{aidan.sedgewick@nbi.ku.dk}}
    \and
    Christa~Gall\orcidGall \inst{1}
    \and
    Luca~Izzo\orcidIzzo \inst{1,2} 
    \and
    Adriano~Agnello\orcidAgnello \inst{1,3}
    \and
    Charlotte~R.~Angus\orcidAngus \inst{1,4}
    \and
    Jens~Hjorth\orcidHjorth \inst{1}
    \and
    Arthur~Kadela\orcidKadela \inst{1}
  }

  \institute{
    DARK, Niels Bohr Institute, University of Copenhagen, Jagtvej 155A, DK-2200 Copenhagen N, Denmark
    \and
    INAF, Osservatorio Astronomico di Capodimonte, Salita Moiariello 16, I-80121 Naples, Italy
    \and 
    STFC Hartree Centre, Sci-Tech Daresbury, Keckwick Lane, Daresbury, Warrington (UK) WA4 4AD
    \and
    Astrophysics Research Centre, School of Mathematics and Physics, Queen’s University Belfast, Belfast BT7 1NN, UK
  }

   \date{Received DDMMYYYY; accepted DDMMYYYY}

 
  \abstract
    {
        The upcoming Vera C. Rubin Legacy Survey of Space and Time (LSST) will discover tens of thousands of astrophysical transients per night, far outpacing available spectroscopic follow-up capabilities.
        Carefully prioritising candidates for follow-up observations will maximise the scientific return from small telescopes with a single-object spectrograph.
    } 
    {
        We introduce AAS2RTO, an astrophysical transient candidate prioritisation tool written in Python. 
        AAS2RTO is flexible in that any number of criteria that consider observed properties of transients can be implemented. 
        The visibility of candidates from a given observing site is also considered.
        The prioritised list of candidates provided by AAS2RTO is continually updated when new transient data are made available.
        Therefore, it can be applied to observing campaigns with a wide variety of scientific motivations.
    }
    {
        AAS2RTO uses a greedy algorithm to prioritise candidates. 
        Candidates are represented by a single numerical value, or `score'. 
        Scores are computed by constructing simple numerical factors which individually consider the competing facets of a candidate which make it suitable for follow-up observation. 
        AAS2RTO is currently configured to work primarily with photometric data from the Zwicky Transient Facility (ZTF), distributed by certified LSST community brokers.
    }
    {
        We provide an example of how AAS2RTO can be used by defining a set of criteria to prioritise observations of type Ia supernovae (SNe Ia) close to peak brightness, in preparation for observations with the spectrograph at the Danish-1.54m telescope.
        Using a sample of archival alerts from ZTF, we evaluate the criteria we have designed to estimate the number of SNe Ia that we will be able to observe with a 1.5m telescope.
        Finally, we evaluate the performance of our criteria when applied to mock LSST observations of SNe Ia.
    }
    {}


   \maketitle
%

\section{Introduction}
\label{sec:intro}
The Vera C.\ Rubin Observatory Legacy Survey of Space and Time \citep[LSST,][]{Ivezic2019} will monitor 
the entire southern sky for the next ten years.
This next generation optical photometric survey will produce a unique data set that will enable to address the most pressing science questions revolving around the formation and evolution of our universe and objects within it.
LSST's special telescope and camera design results in a large field of view of 9.6 square degrees \citep{Graham2020} allowing it to swiftly image large areas of the sky. 
Every night, about 1000 images will be taken in a selection of six filters ($u,g,r,i,z$ and $y$) in a pre-defined pattern.
This together with the high depth of each image (e.g., the anticipated 5$\sigma$ point source $r$-band depth $\approx$ $24.5\text{ mag}$) makes LSST unique in detecting a large range of astrophysical objects and transient phenomena that are either faint or change in either brightness or position on short timescales.
Within 60 seconds of observations, LSST will alert the community of these events, totalling around 10 million alerts every night \citep[][]{Ridgway2014}. 
Therefore, LSST will enable exploration of yet unknown regions in the phase space of transient phenomena \citep{Ivezic2019}, extending the volume-time space to about 100 times over ongoing surveys such as the Young Supernova Experiment \citep{Jones2021, Aleo2023}, the Zwicky Transient Facility \citep[ZTF,][]{Bellm2019a}, the Asteroid Terrestrial-impact Last Alert System \citep[ATLAS,][]{Tonry2018}, BlackGEM \citep{Bloemen2015}, the All-Sky Automated Survey for Supernovae \cite[ASAS-SN,][]{Shappee2014}.
Planned complementary photometric and spectroscopic surveys such as La Silla Schmidt Southern Survey \citep[LS4,][]{Nugent2020}, which is expected to begin operation in 2024/2025, will aid this endeavour.

LSST alerts will be distributed to the astronomical community through dedicated \textit{community alert brokers} (`brokers') that include \textsc{fink} \citep{Moller2021}, the Automatic Learning for the Rapid Classification of Events broker \citep[ALeRCE,][]{Forster2021}, \textit{Lasair} \citep{Smith2019}, and the Arizona-NOIRLab Temporal Analysis and Response to Events System \citep[ANTARES,][]{Matheson2021}. 
All brokers will filter and process the received LSST alerts according to broker-specific criteria. Potential `bogus' (false positive) alerts will be filtered out prior to processing. 
For alerts considered `real', additional information, such as the predicted classifications of an astrophysical object, can be added to the original LSST alert. Brokers refer to this extra information as `annotating'. 
However, all brokers have different approaches to annotating alerts, and distributing alerts in real-time. 
Despite this, the capacity to query archival alerts, and entire lightcurves, is common to all brokers.

With ongoing dedicated photometric transient surveys, such as the Zwicky Transient Facility \citep[ZTF,][]{Bellm2019b} (with single-epoch $5\sigma$-depth of $r=20.6\text{ mag}$), it has been possible to either spectroscopically or photometrically follow-up a large fraction of the newly discovered transient candidates. 
For instance, the ZTF Bright Transient Survey \citep[BTS,][]{Fremling2020, Perley2020}, has obtained spectra of almost every ZTF candidate with peak brightness $r_{\rm peak}\lesssim 18.5\text{ mag}$ and has spectroscopically confirmed over 8800 transients within the past six years\footnote{\url{https://sites.astro.caltech.edu/ztf/bts/bts.php}}. 

However, the unprecedented photometric depth and resulting high rate of alerts (e.g., $\sim$ 1 million supernovae (SNe) per year) from LSST challenges current spectroscopic follow-up capacities. 
To this end, dedicated new instruments and surveys, such as the Time-Domain Extragalactic Survey \citep[TiDES]{2019Msngr.175...58S} at the 4-meter Multi-Object Spectroscopic Telescope are currently being developed.
The Son of X-Shooter spectrograph \citep[SOXS,][]{Schipani2018} for the 3.58-m European Southern Observatory (ESO) New Technology Telescope at La Silla will cover a wavelength range $0.35$ to $2.0\text{ micron}$, with a significant portion of its observing time dedicated to characterising newly-discovered transients.

On the other hand, existing smaller telescopes may be utilised for spectroscopic follow-up observations of specific targets and for specific science cases, aiding the limited spectroscopic resources currently available. 
To maximise the scientific return from small telescopes, careful prioritisation of LSST-discovered transient candidates will be essential. 

Here, we have developed the Automated Alert Streams to Real-Time Observations tool (AAS2RTO). 
The primary goal is to aid in the prioritisation of LSST discovered transient candidates to optimise spectroscopic follow-up observations with the Danish-1.54m telescope (DK1.54m).
There are already several existing tools to aid in prioritising observations for a very large variety of scientific goals \citep[e.g.,][]{SteeleCarter1997, Rana2017, Dyer2018, Dyer2020, Hundertmark2018, Andersen2019}.
However, none exactly meet the requirements for using the DK1.54m as a flexible LSST spectroscopic follow-up resource.

The DK1.54m is located at the European Southern Observatory (ESO) at La Silla, Chile, which is ${\sim}100\text{ km}$ north of the Vera C. Rubin Observatory located at Cerro Pach{\'o}n. 
The main instrument at the DK1.54m is the Danish Faint Object Spectrograph and Camera \citep[DFOSC,][]{Andersen1995}, which can be used for photometry and spectroscopy, covering the same optical to near-infrared wavelength-range as LSST. 
DFOSC has a history of spectroscopic follow-up of a wide variety of transients.
For instance, it was used in the study of SN 1998bw and associated Gamma Ray Burst GRB980425 \citep{Patat2000, Patat2001}, and in identifying the optical counterpart of the X-ray pulsar GS 1843+009 \citep{Israel2001}.
However, since 2003, the spectrograph has been decommissioned. 
The DK1.54m has recently been used solely for photometric observations, for projects such as Ondrejov Asteroid Photometry Project \citep[e.g.,][]{Pravec2014, Pravec2024}, the Microlensing Network for the Detection of Small Terrestrial Exoplanets \citep[MiNDSTEp,][]{BragaRibas2014, Southworth2016, Giannini2017}.
With the upcoming event of LSST and the need for spectroscopy, efforts are being made to re-commission the spectrograph.

Despite AAS2RTO being primarily intended to be used with the DK1.54m, the algorithm is flexible, and therefore can be adjusted for any other telescope.
Further, it can be adapted to a wide range of scientific goals and observing strategies. 
This also requires that different data and alert streams from either private or public surveys can easily be incorporated, which AAS2RTO's modular structure allows.
Here, to test the algorithm, we primarily use data and alerts from ZTF as a substitute for future streams from LSST.

The paper is organised as follows: 
In Sect.\ \ref{sec:concept}, we describe the main concept of AAS2RTO, and outline the structure of the prioritisation algorithm. 
We also outline the key sources of data that are used in this work.
Sect.\ \ref{sec:test_case} describes the implementation of AAS2RTO for an example science case, which is aimed at obtaining spectra of type Ia supernovae at peak brightness.
We use two years of archival data from ZTF to illustrate candidates which would have been highly ranked by AAS2RTO.
Finally, in Sect.\ \ref{sec:comparison}, we compare the prioritisation strategy used by AAS2RTO to a selection of other available schedulers and prioritisation tools.
AAS2RTO is written in \textsc{python} and is publicly available\footnote{AAS2RTO is available at \url{github.com/aidansedgewick/dk154-targets-py38}}.

\section{Methods}\label{sec:concept}

The aim of AAS2RTO is to automatically ingest and process transient data streams, and rank transient candidates according to the scientific interest of a user. 
This said, AAS2RTO is not a broker, as it compiles alert streams that have already been pre-filtered by brokers. 
The main functionality of AAS2RTO is to decide which of these pre-filtered alerts have attributes that match best the criteria of a given scientific use-case. 
Our algorithm neither filters or classifies all transient alerts from a given survey, nor does it redistribute alerts.

AAS2RTO uses a \textit{greedy} strategy \citep[e.g.][]{Black2005, Goodrich2014} for prioritising candidates
Such a strategy is also referred to as a `dispatch' or `just-in-time' strategy by other observation management systems. 
\cite{SteeleCarter1997} summarise this strategy as finding ``the best observation at a given time based on the current telescope state and [observing] conditions without any attempt to look ahead''.
AAS2RTO assesses an unordered set of candidates and computes a single value for each candidate. 
This value quantifies how `interesting' (favourable for observation) the candidate is according to the user-specified criteria.
We refer to this single value as the `score', and it is described in detail in Sect.\ \ref{sec:scoring_function}.
This single score is used to rank candidates, to determine which candidate is best scheduled for the next possible observation. 

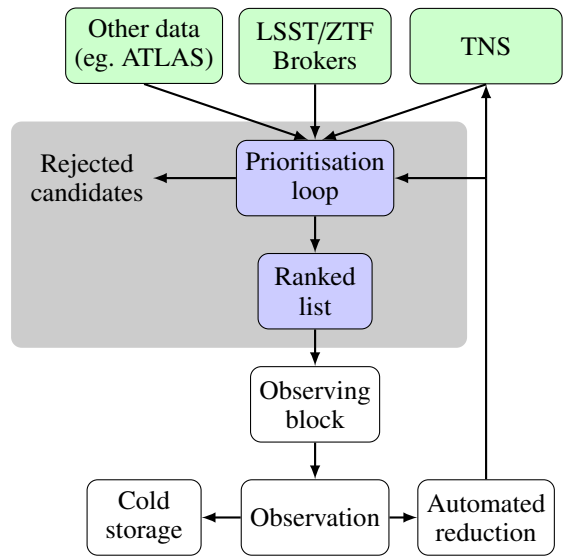
\begin{figure}
    \centering
    \begin{tikzpicture}
        \node(aa2rto) [align=center, rounded corners, minimum width=6cm, minimum height=3cm, xshift=-1cm, yshift=-0.75cm, fill=black!20]{};
    
        \node(priorityloop)[aas2rto, align=center]{Prioritisation\\ loop};

        \node(tns)[data, align=center, above of=priorityloop, xshift=2.25cm, yshift=0.75cm]{TNS};
        \node(brokers)[data, align=center, above of=priorityloop, xshift=0cm, yshift=0.75cm]{LSST/ZTF\\Brokers};
        \node(otherdata)[data, align=center, above of=priorityloop, xshift=-2.25cm, yshift=0.755cm]{Other data\\(eg.\ ATLAS)};
        
        \draw [arrow] (tns.south) -- (priorityloop.80);
        \draw [arrow] (brokers.south) -- (priorityloop.90);
        \draw [arrow] (otherdata.south) -- (priorityloop.100);

        \node(rejects) [align=center, left of=priorityloop, xshift=-2.0cm]{Rejected\\candidates};

        \draw[arrow] (priorityloop.west) -- (rejects.east);

        \node(rankedlist)[aas2rto, align=center, below of=priorityloop, yshift=-0.5cm]{Ranked\\list};

        \draw[arrow] (priorityloop.south) -- (rankedlist.north);

        \node(observingblock)[outofscope, align=center, below of=rankedlist, yshift=-0.5cm]{Observing\\block};

        \draw[arrow] (rankedlist.south) -- (observingblock.north);
        
        \node(observation)[outofscope, align=center, below of=observingblock, yshift=-0.5cm]{Observation};

        \draw[arrow] (observingblock.south) -- (observation.north);

        \node(reduction) [outofscope, align=center, right of=observation, xshift=+1.25cm]{Automated\\reduction};
        \node(storage) [outofscope, left of=observation, align=center, xshift=-1.25cm]{Cold\\storage};

        \node(obs_tns_angle)[right of=observation, xshift=1.25cm]{};
        \node(above_loop)[right of=priorityloop, xshift=1.25cm]{};

        \draw[thick] (reduction.north) -- (above_loop.center);
        \draw[arrow] (above_loop.center) -- (priorityloop.east);
        \draw[arrow] (above_loop.center) -- (tns.south);

        \draw[arrow] (observation.east) -- (reduction.west);
        \draw[arrow] (observation.west) -- (storage.east);   
    \end{tikzpicture}
    \caption{
        An sketch of the logic behind automated observations using AAS2RTO and the Danish-1.54m. 
        Green boxes indicate data compiled by AAS2RTO. 
        White boxes are out of the scope of AAS2RTO.
        The steps made by AAS2RTO in the grey box are detailed in Fig.\ \ref{fig:selector_flowchart}.
    }
\label{fig:obs_flowchart}
\end{figure}

Figure \ref{fig:obs_flowchart} depicts a schematic view of the key components of AAS2RTO in the context of an automated observing cycle with the DK1.54m. 
The primary elements are data acquisition, candidate prioritisation, observing block generation, actual observation, (automated) data reduction, distribution and storage. 
For observations specifically with the DK1.54m, we are separately developing reduction pipelines with PypeIt \citep{Prochaska2020}.
Data can then be made publicly available (for example, via the TNS).

\begin{figure}
    \centering
    \begin{tikzpicture}

        \node(start)[align=center, aas2rto, align=center]{Compile\\data};

        \node(datain)[above of=start, yshift=0.5cm]{Data in};
        \draw[arrow] (datain.south) -- (start.north);
        
        \node(prefilter)[aas2rto, align=center, below of=start, xshift=-0.7cm, yshift=-0.75cm]{Pre-\\filtering};
        \draw[arrow] (start.south) --  node[anchor=east]{new} (prefilter.north);

        \node(existingblank)[below of=start, xshift=1cm, yshift=-0.75cm]{};
        \draw[] (start.south) -- node[anchor=west]{existing} (existingblank.center);

        \node(modeling)[aas2rto, align=center, below of=prefilter, xshift=0.7cm, yshift=-0.75cm]{Build\\models};

        \draw[arrow] (prefilter.south) -- (modeling.north);
        \draw[arrow] (existingblank.center) -- (modeling.north);
        
        \node(firstreject)[align=center, left of=prefilter, xshift=-1.25cm]{Rejected\\candidates};

        \draw[arrow] (prefilter.west) -- (firstreject.east);

        \node(scoring)[aas2rto, align=center, right of=modeling, xshift=1.5cm]{Compute\\ score};

        \draw[arrow] (modeling.east) -- (scoring.west);

        \node(secondreject)[align=center, below of=scoring, yshift=-0.5cm]{Rejected\\ candidates};

        \draw[arrow] (scoring.south) -- (secondreject.north);

        \node(rankedlist)[aas2rto, align=center, right of=start, xshift=1.5cm]{Rank\\candidates};

        \draw[arrow] (scoring.north) -- (rankedlist.south);

        \draw[arrow] (rankedlist.west) -- (start.east);

        \node(output)[align=center, right of=rankedlist, xshift=1.0cm]{Output\\ranked\\lists};

        \draw[arrow] (rankedlist.east) -- (output.west);
        
    \end{tikzpicture}
    \caption{
        The prioritisation loop of AAS2RTO. 
        First, data is compiled (step (\ref{item:data}); Sect.\ \ref{sec:data_sources}) from broadcasting services (e.g. ZTF/LSST brokers) and non-broadcasting services (e.g. ATLAS/TNS). 
        A pre-filtering step (step (\ref{item:init_score})) identifies new candidates which can immediately be removed without detailed modelling or visibility considerations. 
        Theoretical models (e.g. lightcurve fits) are fit to new candidates (step (\ref{item:models})), or candidates which have received new data/alerts in the data compilation step. 
        The full score is computed for each candidate (step \ref{item:full score}; Sect.\ \ref{sec:scoring_function}), considering the lightcurve, and models fit in the previous step, and observing site conditions.
        Finally, list of candidates are produced, in order of descending score (step (\ref{item:ranked_lists}); Sect.\ \ref{sec:ranked_lists})
    }
    \label{fig:selector_flowchart}
\end{figure}
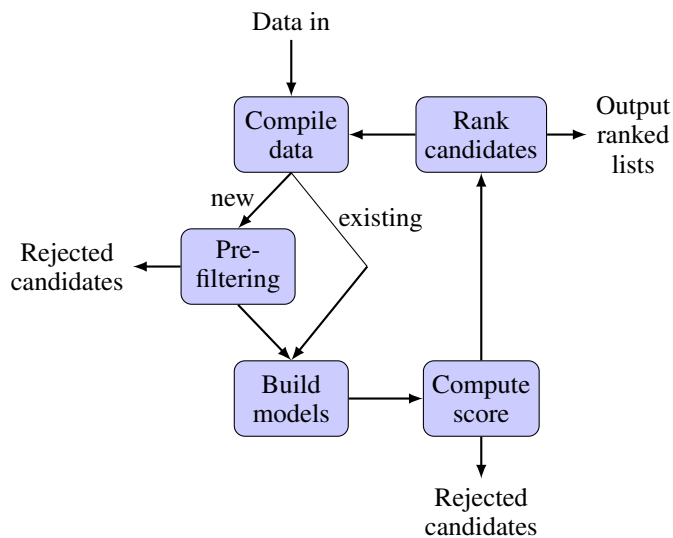

When LSST begins operation, alerts will be produced and distributed at timescales that can be shorter than the required exposure times for a given candidate (particularly for spectroscopy). 
To optimise observations, a ranked candidate list needs to be updated on timescales at least shorter than that of spectroscopic observations, but ideally with a similar frequency to that of the alert distribution. 
AAS2RTO therefore repeatedly performs a `prioritisation loop'.
This loop is visualised in Fig.\ \ref{fig:selector_flowchart} and can be understood as follows:
\begin{enumerate}
    \item \textit{Data acquisition}:
        \label{item:data}
        AAS2RTO receives alerts from broadcasting services (for example, \textsc{fink} or \textit{Lasair}) of new detections of candidates.
        Transient surveys such as LSST or ZTF send `raw alerts' to brokers, which are data packet of a single photometric detection of an astrophysical event that changed in either brightness or position. 
        Along with the photometric information of the detection (aperture magnitudes and 5$\sigma$ detections limits), an alert packet contains information of the celestial coordinates, a timestamp and small image cutouts (single-epoch, static sky and difference image) of the event. 
        Alerts are aggregated (by on-sky position) into \textit{objects} or \textit{candidates}. 
        Brokers then ingest, filter, annotate and redistribute viable alerts (see Sect.\ \ref{sec:intro}). 
        AAS2RTO integrates some alert information (i.e., magnitude, timestamp) either into an existing lightcurve of an already-known candidate, or into photometric data that are queried for new candidates.
            Any additional candidates which users have specified manually can be included at this stage. 
            This allows for Target of Opportunity (ToO) events to be scored and ranked along with candidates which are received from broadcasting services.
            To add new ToOs in AAS2RTO, users must write the details in simple \textsc{ascii} text files which are saved in a pre-defined directory.
        Finally, for each candidate, other lightcurves and additional data from non-broadcasting services are queried.
    \item \textit{Candidate pre-filtering}:
        \label{item:init_score}
        We compute a first estimate of the `score' for any previously-unknown candidate.
        This initial pre-filtering uses the components of the score which do not depend on any models, and therefore are not expensive to compute.
        This is explained more thoroughly with an example in Sect. \ref{sec:score_dk154}.
        Any new candidates which can already be labelled unsuitable in this step (\ref{item:init_score}) will be removed.
        For prioritising observations of transients, the age of the transient (measured by the total duration of the lightcurve, for example) often plays a key role in whether it aligns with the science criteria.
    \item \textit{Fit relevant models}:
        \label{item:models}
        Theoretical models (e.g.\ lightcurve fits for transients) can be useful aids in identifying and prioritising viable candidates.
        Any theoretical models are fit \textit{after} the pre-filtering stage so that computation is not spent on candidates which are already known to be unsuitable.
    \item \textit{Compute full score}:
        \label{item:full_score}
        Taking all compiled information into account (lightcurves, theoretical models), the full `score' for each candidate\label{item:score} can be computed. 
        At this stage, we also consider observing site-specific criteria.
        For instance, we calculate if candidates are actually observable from the observing site of interest (i.e., the DK1.54m at La Silla).
        As observers may have access to more than one observatory (for parallelising observations of many candidates, for example), we compute the site-specific components of the score for each observing site in a pre-defined list provided by users.
    \item \textit{Candidate removal}:
        \label{item:full score}
        We remove candidates with a score which labels them as unsuitable because of their observed characteristics.
        We note that candidates are not removed because of site-specific reasons alone.
        For example, a candidate is not removed for being below the horizon, as it may be observable later, or be currently observable from another observatory.
    \item \textit{Compile ranked lists}:
        \label{item:ranked_lists}
        We compile a list of the remaining candidates, ranked by the score computed in Step (\ref{item:score}) using the user-defined science criteria.
        We produce additional ranked lists for each observing site of interest which exclude candidates that are not visible.
    \item \textit{Inform users}: 
        AAS2RTO optionally sends messages to users which contain candidate properties and lighcurve figures, and visibility plots for the list of observing sites.
        These messages only contain candidates which have been updated in the current iteration of the prioritisation loop.
        These messages are detailed in Appendix \ref{sec:messengers}.
\end{enumerate}

\subsection{Data sources}\label{sec:data_sources}

The transient survey that is closest to the upcoming LSST survey in terms of footprint, cadence and alert distribution is currently ZTF. 
Therefore, we choose ZTF as our baseline survey to develop and test AAS2RTO.
ZTF monitors the entire northern sky visible from Palomar, California (USA), approximately every two nights in two photometric broadband filters: ZTF $g$- and $r$-band, covering a wavelength range of $3676-5613 \AA$ and $5497-7394 \AA$ respectively.

The ZTF camera provides an extremely wide $47\text{ deg}^{2}$ field of view, and reaches median $5\sigma$ limiting magnitudes of $20.8\text{ mag}$ and $20.6\text{ mag}$ in $g$- and $r$-band respectively (in $30\text{ sec}$ exposures).
In its first 2.5 years of operation, it discovered $>3000$ type Ia supernovae, of which 934 have spectra \citep{Dhawan2022}. 
As of 2024, ZTF has spectroscopically confirmed nearly 9000 transients\footnote{\url{https://sites.astro.caltech.edu/ztf/bts/bts.php}}. 
\cite{Bellm2016} developed the survey speed metric $\dot{V}_{-19}$ (measured in $\text{Mpc}^{3} \text{ s}^{-1}$), which measures the comoving volume in which an object of absolute magnitude $-19$ (characteristic of supernovae type Ia) can be detected per exposure time.
By this metric, and by the simpler \'{e}tendue ($A\Omega$, light-collecting area times field of view, $\text{m}^2\text{ deg}^{2}$), ZTF is the fastest existing transient survey to date.
ZTF has $\dot{V}_{-19}=2.5\times10^{4} \text{ Mpc}^{3} \text{ s}^{-1}$ and $A\Omega=53.1\text{m}^2\text{ deg}^{2}$.

To compare, LSST will have $\dot{V}_{-19}=3.7\times10^{5} \text{ Mpc}^{3} \text{ s}^{-1}$ and $A\Omega=319.5\text{ m}^2\text{ deg}^{2}$.
LSST will be the fastest transient survey by an order of magnitude, when it begins operation.
Since all LSST alerts will be distributed by brokers which are currently distributing ZTF alerts (see Sect.~\ref{sec:intro}), we can confidently use these brokers for testing and developing AAS2RTO.
The specific brokers used by AAS2RTO are discussed in more detail in Sect.\ \ref{sec:brokers}.

The Asteroid Terrestrial-impact Last Alert System \citep[ATLAS,][]{Tonry2018} is an astronomical survey which scans the whole sky every two nights.
ATLAS was proposed as a system to detect potentially hazardous near-Earth asteroids, and has discovered 976 such objects since beginning operation in 2015 (as of the end of 2023\footnote{CNEOS discovery statistics: \url{https://cneos.jpl.nasa.gov/stats/site_all.html}}).
Originally operating with two $0.5\text{ m}$ independent units at Haleakal{\=a} Observatory and Mauna Loa Observatory in Hawai{`}i, USA, two more units have since been added at Sutherland Observatory, South Africa, and El Sauce Observatory, Chile.
The survey is conducted in two broad photometric bands: ATLAS-$o$ (``orange'', $420$--$650\text{ nm}$) and ATLAS-$c$ (``cyan'', $560$--$820\text{ nm}$).
A single ATLAS unit has an \'{e}tendue $A\Omega=11.8\text{ m}^2\text{ deg}^{2}$.

As a high-cadence survey, ATLAS is an extremely useful resource for a wide variety of scientific interests in addition to its asteroid detection purpose.
For example, it is a proven resource for variable star discovery and classification having discovered >400,000 candidates in the first ATLAS variable star catalogue \citep{Heinze2018}. 
Furthermore, as of the end of 2023, it was the instrument of discovery for over 3,700 spectroscopically-confirmed SNe Ia\footnote{Classified SNe listed in the Transient Name Server database to 31-12-2023.}. 
In particular, the high cadence of ATLAS data enabled the discovery of the unprecedented `early flux excess' seen in the 02es-like Type-Ia supernova 2022ywc \citep{Sriatstav2023}.

The Transient Name Server\footnote{TNS: \url{https://www.wis-tns.org/}} \citep{GalYam2021} is a database of known transient objects, continuously updated with discoveries from many transient surveys.
For each transient, the TNS database provides various information such as the coordinates, name of the discovery team and instrument, and the time, brightness, and photometric band of the discovery.
For transients which have been observed spectroscopically, the classification spectra and details, such as the astrophysical type and redshift, $z$, are made available.
For instance, there are more than $11,000$ type Ia supernovae with spectroscopically-measured redshifts as of the end of 2023.

\subsection{Data collection and community brokers}\label{sec:brokers}
We are listening for ZTF alerts from the \textsc{fink} broker\footnote{\textsc{fink} API: \url{https://fink-portal.org/api}} \citep{Moller2021}.
\textsc{Fink} annotates alerts with predicted classifications of physical type, using machine learning methods.
ZTF alerts are then broadcast with Apache Kafka in `streams' according to this predicted class.
There are streams of alerts for transients classified as candidate supernovae \citep{Moller2020, Lenoi2022}, microlensing events, kilonovae \citep{Biswas2023}, active galactic nuclei \citep{Russeil2022}, and others.
This allows us (and any other \textsc{fink} user) to only receive alerts which are relevant to our scientific interests.
\textsc{Fink} flags ZTF alerts as `bad quality' if they do not meet certain internal thresholds.
For instance, an alert is flagged if any of the three image cutouts contains known bad pixels, or the aperture and PSF photometry measurements disagree by more than $0.1\text{ mag}$.
Such bad quality alerts are not distributed with Kafka, but the detections are made available through queries for candidate lightcurves.
We still use these data when fitting lightcurves (see Sect.\ \ref{sec:lc_modelling}), however.

In a similar vein, the ALeRCE broker\footnote{ALeRCE broker \url{https://alerce.online/}} also annotates alerts using machine learning-based classifications, based on both the lightcurve and image cutouts (`stamps').
The ALeRCE `lightcurve classifier' \citep{SanchezSaez2021} has a two-level classification approach.
Firstly, the `top-level' classifies candidate sources into periodic, stochastic and transient classes.
Subsequently, the `bottom-level' further resolves each of these broad classifications into sub-classes (e.g., the transient class is further split into SN Ia, SN Ibc, SNII and super-luminous SN).
On the other hand, the ALeRCE `stamp classifier' \citep{CarrascoDavis2021} has five classes: AGN, SNe, Variable Star, asteroid and `bogus' (false detections).
It uses three stamp types associated with an alert, which are the science image (the single-epoch image of source in its environment or host galaxy), reference image (a deep image of the static sky), and difference image (science minus reference, the source without its environment).
The stamp classifier has the advantage of being able to classify candidates after only a single alert, whereas the lightcurve classifier provides more detailed classification with its two-level approach.

We query alerts from ALeRCE by specifying a class from a particular ALeRCE classifier, a classification threshold, and a time interval (e.g.\ all alerts in the last $24\text{ hr}$ classified as SNe candidates by the stamp classifier, with a minimum $80\%$ confidence). 
At present, we query for `SN'-classified alerts from the stamp classifier, and `SNIa'-classified alerts from the transient top-level lightcurve classifier.

The \textit{Lasair} broker\footnote{\textit{Lasair} broker \url{https://lasair-ztf.lsst.ac.uk/}} \citep{Smith2019} annotates ZTF alerts using the \textit{Sherlock} Sky Context software \citep[][Young et al.\ in prep]{Smith2020, Young2023}. 
\textit{Sherlock} crossmatches alerts with existing deep sky catalogues (by sky position) to identify potential host galaxies for transients, known variable stars and AGN, and close bright galactic stars (angular separation).
\textit{Lasair} enables us to filter alerts by defining quality cuts on the lightcurves of candidates and \textit{Sherlock} annotations on the \textit{Lasair} webpages.
For example, it could be required that lightcurves have a minimum number of detections and minimum peak brightness, or a specific \textit{Sherlock} classification, or minimum separation from bright galactic stars.
Alerts which pass these quality cuts are redistributed with Apache Kafka.

ATLAS provides access to forced photometry through an API\footnote{ATLAS forced photometry API: \url{https://fallingstar-data.com/forcedphot/}} \citep{Shingles2021}.
Forced photometry queries are made by providing a pair of RA and Dec coordinates, and a start and end date (in MJD).
Queries for ATLAS forced photometry return a lightcurve, which is produced by measuring PSF fluxes and magnitudes at the requested coordinates in all difference images available in the requested MJD interval.
However, ATLAS takes four images ($30\text{ sec}$ exposures) of the same field in a 90 minute interval, instead of a single image with a long exposure.
This strategy enables measurements of the proper motion of near-earth and solar system objects, but, the observed flux of extragalactic transients will not change detectably within these 90 minute intervals (within ATLAS flux uncertainties). 
Therefore, we choose to compute the uncertainty-weighted mean of the four forced PSF flux measurements from each 90 minute interval, and convert this mean PSF flux into a magnitude.
Each of the four ATLAS images has an associated $5\sigma$ detection limit, which quantifies the minimum measured flux for which we can consider a detection to be reliable.
We compute the mean $5\sigma$ flux of these four detection limits, and convert it to a magnitude.
The computed mean PSF magnitude is a `valid' detection if it is brighter than the mean $5\sigma$ detection limit.
Otherwise, it is flagged as an upper limit (non-detection).

As the ATLAS server can only process a limited number of forced photometry queries per day, we limit the number of queries queued at any given time. 
ATLAS photometry is only requested for candidates which have already had a valid score computed (see Sect.\ \ref{sec:scoring_function}).
Additionally, we request forced photometry for candidates in descending order of their last computed score. 
In this way, the ATLAS photometry queries which are submitted first are for the candidates which are already highly ranked.

The TNS also provides access to data through an API.
We query for all spectroscopically confirmed transients in the last 30 days and crossmatch the results with the set of all AAS2RTO candidates.
For example, TNS queries could be used to aid an observing campaign which has the aim of observing currently-unclassified transients.
Candidates which have a TNS match could be demoted or rejected.

\subsection{Scores and scoring functions}\label{sec:scoring_function}

The `score', and the `scoring function' which produces the score, are at the heart of AAS2RTO's use as a candidate prioritisation tool.
The primary intent of introducing a scoring function is to balance competing attributes of all available candidates which make a candidate favourable to be observed.
Each of these attributes are considered independently to produce a numerical `factor', $x_{i}$, for each attribute.
The final score, $S$, used for ranking is then computed as the product of all of the factors, and a `base score', $S_{\rm base}$,
\begin{equation}
    \label{eq:score}
    S=S_{\rm base}\,\Pi x_{i}.
\end{equation}
This single score produced for each candidate allows a set of candidates to be ranked, with the candidate with the highest score being ranked first, and so on.
The base score, $S_{\rm base}$, is a value which is fixed for each candidate, and is by default the same for each candidate (e.g.\ $S_{\rm base}=1$).
We provide an example of constructing factors for an example science case in Sect.\ \ref{sec:score_dk154}.

The scoring function also serves as the means of deferring observations of currently unsuitable candidates to later times (`excluding' candidates), or permanently removing candidates from consideration (`rejecting' candidates).
If any of the factors $x_{i}$ indicate that a candidate is irrelevant or unsuitable, the computed score reflects this.
AAS2RTO will reject candidates which have a non-finite score, and exclude candidates which have a negative score.
Importantly, this means that care must be taken in designing factors $x_{i}$ to be positive and finite.
An odd number of factors which are negative will produce a negative score, unintentionally excluding candidates.
Factors which are non-finite can also multiply together to give a non-finite score, causing a candidate to be unintentionally rejected.

Factors $x_{i}$ can be designed to weight attributes which are more important to a certain science case.
For instance, if the goal is to observe bright supernovae, but with a preference towards those which are blue in colour, then two factors $x_{\rm flux}$ and $x_{\rm colour}$ can be constructed to range between $1<x_{\rm flux}<100$ and $1<x_{\rm colour}<10$ (for the range of observed candidate fluxes and colours).
This way, the final score will be more sensitive to changes in brightness than colour.

Modifying the base score is useful if candidates are known ahead of time to be of particular interest.
For example, if there is a new, high-priority candidate which a user wishes to add to the set of existing candidates, it can be added with a larger base score than the default value.
In the example above with $1<x_{\rm flux}<100$ and $1<x_{\rm colour}<10$ and a default value of $S_{\rm base}=1$, the maximum value of the score is $S=1000$. 
Therefore, a high-priority candidate added with $S_{\rm base}=1000$ is guaranteed to be the highest-ranked.
    This is useful for manually including ToOs.
    Candidates added as described above will naturally appear on top of the ranked list. 
    Therefore, remaining observations can continue to be scheduled in the prioritised order that AAS2RTO produces (see Sect.\ \ref{sec:ranked_lists}).

We stress that these scores are used only for ranking candidates, and they do not have a physical interpretation.
It is not the actual value of the score of any given candidate that is important, but its value relative to another candidate.

\subsection{Ranked lists and scheduling}\label{sec:ranked_lists}

The ultimate aim of AAS2RTO is to aid in real-time prioritisation candidates for rapid follow-up observations.
As described in Sect.\ \ref{sec:scoring_function}, the score of each candidate considers any observed quantity which makes it `interesting'.
After computing a score for each candidate, a simple observing schedule can be written by listing the candidates in order of decreasing score.

The visibility of candidates from a given observing site will also have an impact on the priority with which they are observed.
In some cases, candidates may not be at all visible from an observing site.
Scoring functions can take into account observing sites, and therefore, factors $x_{i}$ can be constructed to account for visibility.
This is demonstrated explicitly in Sect.\ \ref{sec:score_dk154}.

Observing campaigns are often carried out using more than one telescope or observatory (for example, to divide candidates amongst several sites).
Therefore, AAS2RTO is capable of considering many observatories of interest to a user.
A ranked list is produced for each observatory, which accounts for candidate visibility (current altitude).
    However, we note that these ranked lists are not a `joint' schedule, and so will not provide a globally optimised observing schedule for a telescope network.
    This type of schedule could be produced using integer linear programming (ILP), similar to the solution demonstrated in \citet{Solar2016} for the Atacama Large Millimeter/submillimeter Array (ALMA) observatory.
Candidates which are not visible from a particular observing site are excluded from the ranked list for that observatory.
The ranked list for a particular observatory can then be used as an observing schedule specific to that site.
    We highlight that the ranked list is updated every iteration of the prioritisation loop (see Fig.\ \ref{fig:selector_flowchart}).

\section{Example science case: selecting SNe Ia at peak brightness}\label{sec:test_case}

As an example instance of AAS2RTO, we will prioritise type Ia supernovae (SNe Ia) which are at their peak brightness, with the aim of taking spectra at this peak.
The most strict criterion for observations is a faint limit of $i<18.5\text{ mag}$.

Although this science case is motivated by the capabilities of the DK1.54m and its proximity to Rubin, the factors we describe here can be applied to any facility (or applied with minor modifications). 

\subsection{Lightcurve fitting}\label{sec:lc_modelling}

The data sources described in Sect.\ \ref{sec:data_sources} provide candidate lightcurves.
However, for optimised observations at a given point in a transient's evolution we need to be able to predict this point of evolution. 
This can be done by fitting the available data that so far have been ingested with either simple fitting functions or physically-motivated lightcurve models. 
Here, for our test science case of predicting the time of peak brightness of an SN Ia lightcurve, we implement two options.
We note that AAS2RTO is not limited to the models we describe in this Section.

We use the Spectral Adaptive Lightcurve Template \citep[SALT,][]{Guy2005, Guy2007} within the \textsc{sncosmo} framework \citep{Barbary2016} to model the lightcurves of candidates.
SALT is an empirical model for describing the evolution of SNe Ia as a function of time.
Specifically, we use the SALT2 revision of the SALT models presented in \cite{Taylor2021}.

The spectral flux density, $F\left(p,\lambda\right)$ of a source computed with the SALT2 model is given by
\begin{equation}
    F\left(p,\lambda\right) = x_{0}\left[M_{0}\left(p,\lambda\right) + x_{1}M_{1}\left(p,\lambda\right)\right]\times10^{-0.4CL\left(\lambda\right)c}
    \label{eq:salt_flux}.
\end{equation}
The model has component templates $M_{0}\left(p, \lambda\right)$, $M_{1}\left(p, \lambda\right)$ and $CL\left(\lambda\right)$, where $p$ is phase \citep[defined for SALT as the time since the peak brightness in the $B$-band,][]{Guy2005}, $\lambda$ is wavelength, and $x_{0}$, $x_{1}$ and $c$ are free parameters.
The three components $M_{0}$, $M_{1}$ and $CL$ have been computed from a sample of 420 SNe Ia, of which 83 have at least one spectrum.
They represent the global model of SNe Ia lightcurves as a function of time and wavelength.
\cite{Guy2007} state that SALT would be equivalent to a principal component analysis if it were not modulated by the colour law $CL\left(\lambda\right)$.
$M_{0}\left(p, \lambda\right)$ and $M_{1}\left(p, \lambda\right)$ are time-dependent model components.
$M_{0}$ and $M_{1}$ respectively encode the mean spectral energy distribution of SNe Ia, and variation from this mean.

The free parameters amplitude $x_{0}$, stretch $x_{1}$ and colour $c$ are determined for a particular supernovae lightcurve, using a least squares fitting process or Monte Carlo methods.
The amplitude parameter $x_{0}$ is related to $m_{B}$, the apparent $B$-band peak magnitude, as $m_{B}=-2.5\log10\left(x_{0}\right)+10.5$ \citep{Kenworthy2021}.
The zero-phase parameter, $t_{0}$ is defined as the time of the peak brightness in the $B$-band ($p=0$, in days, often given as a Modified Julian Date).
This parameter, along with the redshift, $z$, of the source are also allowed to vary. 
We use the dustmaps from \cite{Schelgel1998} to compute the Milky Way dust extinction.
We do not attempt to fit SALT models to lightcurves until there are at least five detections.

As described earlier, we crossmatch all candidates with the recent TNS database entries.
If a candidate has a match from the TNS database with a spectroscopic redshift, we fix the redshift of the SALT model to this value.

An alternative would be to use a functional form, such as the the one suggested in \cite{Bazin2011} for modelling the flux $F^{k}\left(t\right)$:
\begin{equation}\label{eq::bazin}
    F^{k}\left(t\right)=A^{k}\frac 
    {\exp{\left[-\left(t-t_{0}^{k}\right)\middle/\tau_{{\rm fall}}^{k}\right]}} 
    {1+\exp{\left[-\left(t-t_{0}^{k}\right)\middle/\tau_{{\rm rise}}^{k}\right]}} + q^{k},
\end{equation}
where index $k$ is the photometric band (e.g.\ ZTF $g$ or $r$) that the best-fitting parameters $A$, $t_{0}$, $\tau_{\rm fall}$, $\tau_{\rm rise}$ and $q$ are estimated for each band $k$.
$A$ is the amplitude, $\tau_{\rm fall}$ and $\tau_{\rm rise}$ are characteristic fall and rise timescales (respectively), and $q$ is a constant offset.
As derived in \cite{Bazin2011}, for each photometric band $k$, $t_{0}$ is related to the time of maximum $t_{\rm max}$ as 
\begin{equation}
    t^{k}_{\rm max}=t^{k}_{0} + \tau^{k}_{\rm rise} \ln \left(\tau^{k}_{\rm fall}\middle/\tau^{k}_{\rm rise}-1\right),
\end{equation}
meaning that $\tau_{\rm fall} > \tau_{\rm rise}$ is required for a Bazin lightcurve model to have a maximum value.

By default, there are no relationships encoded between free parameters as a function of index $k$ (photometric band), meaning that the parameters for each photometric band are independent.
This is a disadvantage compared with the SALT models, as SALT uses all available detections to find a single set of best-fitting parameters.

As the SALT $M_{0}$ and $M_{1}$ templates are derived from SNe Ia observations, they are not strictly appropriate for modelling other types of SNe.
Nevertheless, we use the SALT templates to fit the lightcurves of all of our candidates, and use them to estimate the time of peak brightness because we will not know in advance the spectral type of the candidate.
Although it is a more general lightcurve model, the Bazin form is less useful for estimating a peak time for a supernovae which is still rising (increasing in brightness), as the $\tau_{\rm fall}$ parameter requires appropriate priors.
As the SALT models are based on templates of SNe Ia lightcurves (which rise and fall), they have a peak `built in'.

AAS2RTO is not limited to SALT or Bazin models, however, and is flexible enough to be adapted to other models that users may prefer.

\subsection{Designing a scoring function}\label{sec:score_dk154}

Here we will describe factors which are used to promote candidate SNe Ia which are close to peak brightness.

\subsubsection{Candidate properties}

The strictest criterion for selecting candidates for the DK1.54m telescope is the limiting magnitude of $i>18.5\text{ mag}$.
Aside from this practical limitation, the example science case does not depend on the observed SNe Ia magnitude, so we promote candidates which are brighter simply because they can be observed with a shorter integration time for a desired signal-to-noise ratio.
We therefore define the factor
\begin{equation}
    x_{\rm mag} = 10^{ 0.5 \times \left(18.5 - m\right)},
\end{equation}
where $m$ is the latest ZTF detection (in mag), in either the ZTF $g$- or $r$-band.
This prescription means that a candidate with $m=16.5\text{ mag}$ will have $x_{\rm mag}=10$.
Candidates with $m>18.5\text{ mag}$ are flagged so that their final score will be negative, so they are excluded from ranked lists.
These candidates are not rejected outright, as they could be rising and have $m<18.5\text{ mag}$ in the future.

To increase the probability that a candidate is real, we also require there to be at least four detections in the lightcurve (from both $g$- or $r$-bands).
Candidates with fewer than this are excluded from ranked lists, meaning that very young candidates are likely to be excluded.
This could be avoided by reducing the minimum number of detections, and using some other criteria to ensure `real' candidates.
For instance, by requiring all detections to meet a prescribed minimum broker classification threshold.
However, as young candidates are not the focus of this example, we do not make this adjustment.

To promote candidates which are near peak brightness, we construct a factor based on a normal distribution with centred at the latest estimate of the SALT zero-phase parameter $t_{0}$ (described in Sect.\ \ref{sec:lc_modelling}), with width $\sigma$,
\begin{equation}\label{eq:time_to_peak}
    x_{\rm peak} = A \times \exp\left[-\frac{\left(t_{\rm obs}-t_{0}\right)^{2}}{2\sigma^{2}}\right].
\end{equation}
Here, $t_{\rm obs}$ is the time of the next observation.
We choose the width of the distribution $\sigma=1\text{ day}$.
This can be adjusted depending on how critical it is to select candidates at their peak brightness. 
For instance, if spectra within two days of the peak are acceptable, the width of the function could be adjusted to $\sigma=2\text{ days}$.
We set the amplitude parameter $A=30$, so that the factor $x_{\rm peak}$ has a maximum value of 30 (when $t_{\rm obs}=t_{0}$).
This choice of maximum value gives more weight to candidates which are close to the peak than the maximum expected values of $x_{\rm mag}\sim10$.
We also set a minimum value of $x_{\rm peak}=10^{-2}$, which occurs around when $|t_{\rm obs}-t_{0}|> 4$.
If the SALT model fitting fails, we set $x_{\rm peak}=1.0$.

As it is based on a function which is symmetric about $t_{0}$, the factor $x_{\rm peak}$ alone gives equal priority to a candidate two days before $t_{0}$ and a candidate two days after $t_{0}$. 
It is better to promote candidates before peak brightness rather than after (i.e., still increasing in brightness) - it is still possible to observe these candidates at their peak brightness.
We define factors, $x_{\rm rise}^{k}$ (for each photometric band $k$ in the lightcurve), which are the fraction of detections in a lightcurve which are brighter than the previous one.
That is,
\begin{equation}\label{eq:x_rise}
    x_{\rm rise}^{k} = \sum_{i=1}^{N^{k}-1} 
        \left. \left[m^{k}_{i+1} < m^{k}_{i}\right]\ \middle/\  {\left(N^{k}-1\right)}\right.,
\end{equation}
where index $k$ is the photometric band (ZTF $g$ or $r$, ATLAS $o$ or $c$), and $m^{k}_{i}$ is the magnitude of the $i^{\rm th}$ detection in band $k$ (in mag).
$N^{k}$ are the number of detections in band $k$. 
Iverson bracket notation evaluates to 1, if the logical statement enclosed is true, and 0 otherwise.
Here, this means $\left[m^{k}_{i+1} < m^{k}_{i}\right]=1$ if $m^{k}_{i+1} < m^{k}_{i}$ is true, and 0 otherwise.
The denominator in Eq.\ \ref{eq:x_rise} is $N^{k}-1$ as this is the number of consecutive pairs from $N^{k}$ detections.
This factor can vary between 0 and 1, and we choose to reject candidates if all $x_{\rm rise}^{k} < 0.4$ and $N^{k}>2$.
If $N^{k}\leq3$, we choose $x_{\rm rise}^{k}=1.0$.
Candidates are therefore only rejected due to $x_{\rm rise}^{k}$ if there are at least four detections in each band.
In this way we avoid rejecting candidates with a pair of detections, which happen to appear to be declining in brightness only due to one poor photometric detection.

We also consider the time since the first observation, $T$ (in days), again with the motivation of disfavouring candidates which are past the peak brightness.
We define
\begin{equation}
    x_{\rm span}=\begin{cases}
        1 & T<20 \text{ days};\\
        L(T;r,x_{\rm m}) & \text{otherwise},\\
    \end{cases}
\end{equation}
with logistic function, $L$, as: 
\begin{equation}
    \label{eq:logistic}
    L\left(x;r,x_{\rm m}\right)=\frac{1}{1+\exp\left(-r\left(x-x_{\rm m}\right)\right)}.
\end{equation}
The logistic function $L$ varies smoothly from 0 to 1 around the midpoint parameter, $x_{\rm m}$. 
The steepness of the logistic is set by the parameter $r$, with larger values of $r$ producing steeper increase.
Negative values of $r$ mean that $L$ instead starts at 1 for $x<x_{\rm m}$ and decreases to 0.
We choose $x_{\rm m}=25\text{ days}$ and $r=-1\text{ days}^{-1}$, which means that the function decreases from 0.99 to 0.01 between 20 and 30 days, and 0.9 to 0.1 between 23 and 27 days.
Finally, we reject candidates with $T > 30\text{ days}$.
We have designed this factor to decrease after 20 days because the rise time of SNe Ia is around 19 days in the rest-frame \citep[e.g.][]{Reiss1999, Firth2015}.

As mentioned above, the factors $x_{\rm rise}$ and $x_{\rm span}$ serve to promote supernovae which have not yet passed peak brightness, although in a less direct way than comparison with the best estimate of the time of the peak.
However, this is useful in cases where SALT model fitting fails for a candidate.

We then compute the score for each candidate SNe Ia following Eq.\ \ref{eq:score},
\begin{equation}
    \label{eq:score_sneia}
    S_{\rm Ia} = 
    S_{\rm base}\, x_{\rm mag}\, x_{\rm peak}\, x_{\rm rise}\, x_{\rm span}, 
\end{equation}
and we choose the base score $S_{\rm base}=1$.

In Sect. \ref{sec:concept}, in step (\ref{item:init_score}) of the numerical description, we describe using the score for an initial check, which does not depend on any of the factors dependent on any model.
Here, those are the three factors ${\rm mag}$,  $x_{\rm rise}$ and $x_{\rm span}$.
Candidates which are rejected because of any of these factors should not have SALT models fit.

\subsubsection{Candidate visibility}\label{sec:candidate_visibility}

So far all of the factors we have described consider only the observed properties of a candidate.
We also consider the visibility of candidates from a given observatory.
A simple prescription uses the current airmass, $X$, or altitude, $a$, of the candidate
\begin{equation}
    x_{\rm{alt}}=\frac{1}{X}\sim\sin\left({a}\right),
    \label{eq:altitude}
\end{equation}
using the simple secant approximation $X\sim{\rm sec}\left({90^{\circ}-a}\right)$, where $a$ is the altitude of the candidate above the horizon in degrees.
This is a good approximation for $a \gtrsim 5^\circ$ \citep{YoungIrvine1967, KastenYoung1989}. 
This definition of $x_{\rm alt}$ means $0<x_{\rm alt}<1$.
We choose to exclude candidates with $a<30^{\circ}$ (where airmass $X>2$).

\begin{figure}
    \centering
    \includegraphics{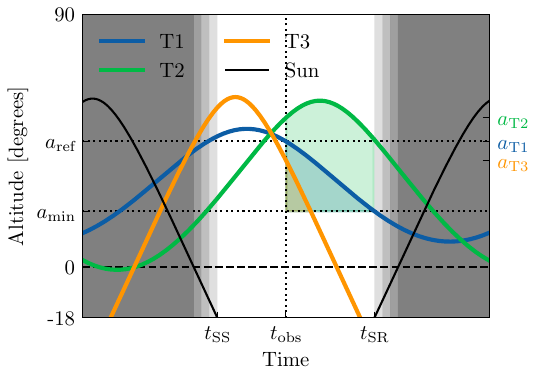}
    \caption{
        The altitude of three hypothetical candidates (T1, T2, T3) from La Silla on the spring equinox.
        The current time $t_{\rm obs}$ is indicated, and $t_{\rm SS}$ and $t_{\rm SR}$ are the sunset and sunrise time respectively.
        The shaded grey regions are day and twilight (the Sun's altitude $a>-18^{\circ}$, when astronomical observations are not possible).
        The shaded region under each altitude curve between $t_{\rm obs}$ and $t_{\rm SR}$ is $A_{\rm vis}$, computed in Eq.\ \ref{eq:A_vis}.
        The reference altitude $a_{\rm ref}$ for normalisation is also indicated.
    }
    \label{fig:altitude}
\end{figure}

An alternative factor considers how visible a candidate will be for the remainder of the night.
We first define the quantity $A_{\rm vis}$ by considering the integral of a candidate's altitude as a function of time, $a\left(t\right)$,
\begin{equation}
    A_{\rm vis} = 
            \int_{t_{\rm obs}}^{t_{\rm SR}} \bigl(a\left(t\right)-a_{\rm min}\bigr)\ dt,
    \label{eq:A_vis}
\end{equation}
where $t_{\rm obs}$ is the time of the next observation (when the score will be evaluated) and $t_{\rm SR}$ is the time of the following sunrise (i.e., the end of the night's observations).
If the current time is during the daytime, we set $t_{\rm obs}=t_{\rm SS}$, the time of the upcoming sunset (so $t_{\rm SS}\leq t_{\rm obs} <t_{\rm SR}$).
Altitude $a_{\rm min}$ is the minimum acceptable altitude that a candidate can be observed at.

The factor $x_{\rm vis}$ is then
\begin{equation}
    x_{\rm vis} = \left(
        \frac{A_{\rm vis}}{
            (a_{\rm ref}-a_{\rm min})(t_{\rm SR}-t_{\rm obs})
        }
    \right)^{-1},
    \label{eq:x_vis}
\end{equation}
with reference altitude $a_{\rm ref}$ used for normalisation (in the numerator). 
A smaller $A_{\rm vis}$ gives larger $x_{\rm vis}$, so that candidates which are going to set below the horizon sooner are promoted.
The numerator is the rectangle enclosed between $t_{\rm obs}$, $t_{\rm SR}$, $a_{\rm min}$ and $a_{\rm ref}$.
We choose $a_{\rm ref}=90^{\circ}$.
This normalisation ensures that the value of $x_{\rm vis}$ does not change dramatically through the night due to decreasing observing time remaining.
Even for $a(t)$ constant (such as Polaris), $A_{\rm vis}$ approaches zero as $t_{\rm obs}$ approaches $t_{\rm SR}$.
Normalisation also makes $x_{\rm vis}$ a dimensionless quantity.

The motivation for factor $x_{\rm vis}$ is illustrated in Fig.\ \ref{fig:altitude}, which shows a hypothetical scenario where there are three candidates to be observed (T1, T2, and T3).
If the goal is to maximise the number of candidates which are observed, candidate T3 should be observed first (as it will set soonest), followed by T1, and finally T2.
Using the simpler approach of $x_{\rm alt}$, Eq.\ \ref{eq:altitude}, the candidates with the current highest altitude would have been observed first.
Here, T2 would have been observed first, followed by T1 - after which time T3 would no longer observable.

One issue with the factor $x_{\rm vis}$ is that it does not have an upper bound for candidates which are very close to setting below $a_{\rm min}$.
This is undesirable if all the other factors for a candidate have been carefully constructed to be well-behaved to reflect the scientific aims of a use.
This could be avoided by modifying the definition to `suppress` extreme values with $\min\left(x_{\rm vis}, A\right)$ function (where $A$ is the maximum allowed), for example.
There is more discussion of the behaviour of $x_{\rm vis}$ in Appendix \ref{sec:vis_factor_appendix}.

For the DK1.54m observatory, we compute for each candidate
\begin{equation}
    \label{eq:score_dk154}
    S_{\rm DK154} = S_{\rm Ia}\, x_{\rm vis},
\end{equation}
where $x_{\rm vis}$ is computed considering the candidate altitude from La Silla observatory.

For the example score we have presented here, we have not implemented a factor to account for the telescope slew time (from one candidate to the next).
Ideally, telescope slew time should be minimised, so we would construct a factor to promote candidates which are near the current telescope pointing.
However, this would require information about the current pointing of the telescope under consideration (for instance, the DK1.54m).
Accessing this information will require specific implementations for each telescope.
For similar reasons, we have not yet included factors to account for observing conditions, such as wind speed.
However, the DK1.54m has strict wind speed limits for safe operation.
Further, we have not yet implemented a factor to consider moon separation or moon phase.

We stress that the absolute value of a candidates' score is unimportant, as it is only used in comparison with other candidates.

\subsection{Testing with archival ZTF data}\label{sec:ztf_archive}

\begin{figure*}[!htp]
    \centering
    \includegraphics{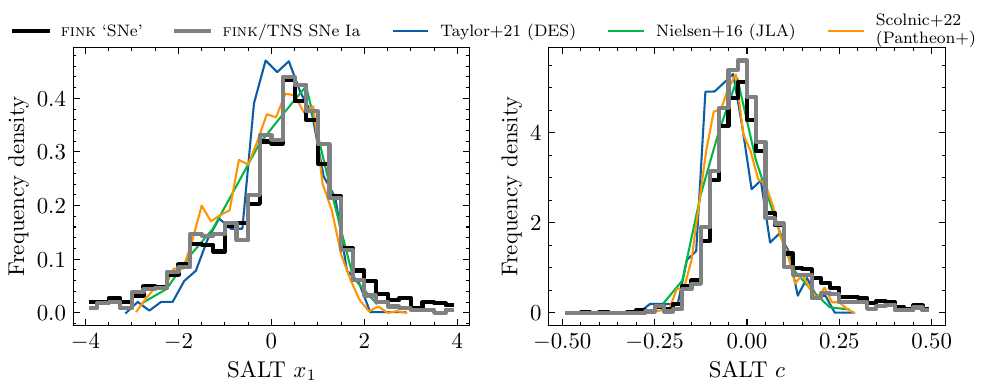}
    \caption{
        The distribution of SALT2 stretch parameter $x_{1}$ (left panel) and colour parameter $c$ (right panel) for all lightcurves with eight or more detections, for the sample of 2205 lightcurves which meet our criteria (black), and for the subset of 1377 which are spectroscopically-confirmed SNe Ia in TNS (grey).
        Distributions of $x_{1}$ (left panel) and $c$ are also shown from \cite{Taylor2021} using the Dark Energy Survey supernovae sample (blue line; taken from their Fig.\ 9), from \cite{Nielsen2016} using the Joint Lightcurve Analysis sample (green; taken from their Fig.\ 1), and from \cite{Scolnic2022} using the Pantheon+ sample (orange).}
    \label{fig:salt_dist}
\end{figure*}

Between the dates 01-October-2020 and 30-October-2022, we select all ZTF lightcurves which have any detection tagged by \textsc{fink} as `SN candidate' or `Early SN Ia candidate'.
There are 314,265 alerts in the ZTF $g$- and $r$-bands, comprising 120,212 unique candidates.
We use this data to illustrate how AAS2RTO might operate with real data from ZTF for the scientific objective outlined above.

We fit SALT2 models to all of the lightcurves with eight or more detections, which have a ZTF $g$-band detection brighter than $g<19\text{ mag}$.
We use \textsc{sncosmo}'s least squares fitting method.
There are 6714 such lightcurves in the sample.
To ensure that the best-fitting SALT models are self-consistent, we require that there are at least four detections after the best-fitting $t_{0}$, and further that at least one of these detections is a minimum of five days after $t_{0}$.
Similarly we require that there are at least three detections before the zero-phase parameter $t_{0}$.
We also remove lightcurves that are longer than $120\text{ days}$ and those which have a gap in the lightcurve longer than $20\text{ days}$.
This cut quickly removes variable or stochastic sources which have been misclassified as SNe, although it may remove `real' SNe which have long lightcurves.
Finally, we require that the best-fitting SALT model has a reduced $\chi^{2}$ of $\chi_{\nu}^{2}<5.0$ ($\chi_{\nu}^{2}$ is $\chi^{2}$ per degree of freedom).
After these quality cuts, there are 2205 candidates remaining.
The distribution of the best-fitting SALT2 stretch $x_{1}$ and colour $c$ parameters of the remaining 2205 lightcurves are shown in Fig.\ \ref{fig:salt_dist} (solid black, labelled as \textsc{fink} `SNe').
For comparison, we also show the distributions of the $x_{1}$ and $c$ parameters from \cite{Taylor2021}, who use the Dark Energy Survey SNe sample, the distributions from \cite{Nielsen2016} using the Joint Lightcurve Analysis sample of SN Ia, and the distributions from \cite{Brout2022} using the Pantheon+ sample\footnote{\url{https://github.com/PantheonPlusSH0ES/DataRelease/}}.
Our distributions of both stretch $x_{1}$ and colour $c$ parameters have heavier tails than the literature distributions, but are still in good agreement.

\begin{figure}
    \centering
    \includegraphics{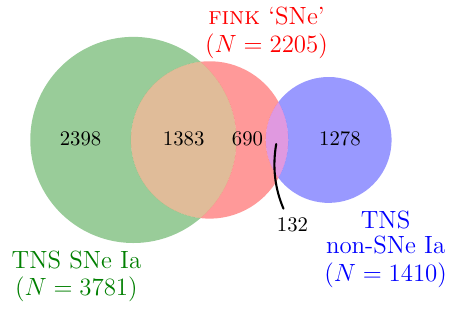}
    \caption{Our sample of SNe Ia selected from \textsc{fink} (central circle, red), and the overlap with the TNS spectroscopically confirmed sample of SNe Ia (left circle, green), and SNe non-Ia (right circle, blue).}
    \label{fig:venn}
\end{figure}

We crossmatch all 2205 candidates with the TNS database of spectroscopically-confirmed SNe (within $5\text{ arcsec}$) which were listed in TNS between 01-October-2020 and 30-October-2022 (the same dates as the \textsc{fink} alert sample).
Of the 5191 spectroscopically confirmed TNS SNe, there are 1515 matches with our sample ($68\%$ of our sample has been spectroscopically confirmed as SNe).
Out of the 3781 SNe Ia in the full TNS dataset, there are 1383 which match with our sample.
By contrast, there are only 132 matches with spectroscopically-confirmed SN which are not type Ia, out of 1410 non-Ias listed in the full TNS dataset (1016 of which are type II SNe).
The $x_{1}$ and $c$ parameter distributions of the spectroscopically-confirmed TNS subset of SNe Ia are also shown in Fig.\ \ref{fig:salt_dist} (labelled as \textsc{fink}/TNS SNe Ia).
These statistics are summarised in Fig.\ \ref{fig:venn}.
The distributions of SALT $x_{1}$ and $c$ for the TNS-confirmed subset of SNe Ia are in good agreement with the full sample of 2205 SNe (solid grey in Fig.\ \ref{fig:salt_dist}). 
This indicates that the \textsc{fink} sample we select and the quality cuts we impose are useful for selecting SNe Ia.

\subsubsection{Timing of type Ia supernovae peak brightness}\label{sec:peak_timing}

Here, we investigate how well we are able to estimate the zero-phase parameter $t_{0}$ as a supernova lightcurve evolves.
We note that the estimates of the zero-phase parameter $t_{0}$ for a still-evolving lightcurve are not intended to be precise or final measurements, and are not used for any cosmological or astrophysical analyses.
Rather, they are a useful tool to aid in prioritisation, and we aim here (in Sect.\ \ref{sec:peak_timing}) to quantify how useful they are.

\begin{figure}
    \centering
    \includegraphics{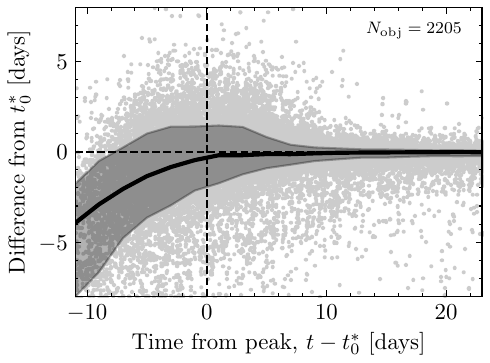}
    \caption{
        The convergence SALT2 zero-phase parameter $t_{0}$ to the final `best' estimate, $t_{0}^{\ast}$, as a function of SN phase. 
        We compute the difference in days between the `latest' value of $t_{0}$ (the value using the first $N$ detections in $g$ and $r$) and $t_{0}^{\ast}$.
        The horizontal dashed line is where $t_{0}=t_{0}^{\ast}$.
        Points about this line are \textit{overestimates} of $t_{0}$ in comparison to $t_{0}^{\ast}$.
        We plot the median $t_{0}-t_{0}^{\ast}$ and $68\%$ confidence interval (solid black line and shaded region), estimated in bins of two days, which is the average ZTF cadence.
    }
    \label{fig:sn_timing}
\end{figure}

Fig.\ \ref{fig:sn_timing} shows how the estimate of $t_{0}$ converges through the duration of a lightcurve.
We use the sample of 2205 SNe Ia that meet the quality cuts we describe above.
For each candidate, we use the SALT2 templates described in Sect.\ \ref{sec:lc_modelling} to model the full available lightcurve to recover the best-fitting parameters, and label the best-fitting zero-phase parameter as $t_{0}^{\ast}$.
We then fit a SALT2 model to the first $N$ detections in the lightcurve and recover the same parameters, and repeat this for each of the $N$ detections in the lightcurve.
That is, if there are 20 detections in a lightcurve, there will be 20 sets of parameters - using all 20 detections, the first 19 detections, the first 18 detections, and so on.

For each value of $t_{0}$ (estimated with increasing number of detections), we compute the difference from the best estimate, $t_{0}-t_{0}^{\ast}$.
These difference values are plotted as a function of $t-t_{0}^{\ast}$ (that is, lightcurve phase, the time since $t_{0}^{\ast}$).
In principle a single candidate could appear $N-1$ times in Fig.\ \ref{fig:sn_timing}, once for each estimate of $t_{0}$, excluding the final estimate (as the difference here is by definition zero).
However, we do not use estimates of $t_{0}$ which use only the first one, two, three or four detections, as we do not attempt to fit SALT models until there are at least five detections.
This is the number of free parameters for a SALT lightcurve fit.

We choose to show this convergence as a function of phase rather than number of detections, as the sampling of lightcurves is different for each object (that is, the time between detections in each lightcurve is not the same).
Additionally, this is how we will use the SALT models in practice, where we compute the factor $x_{\rm peak}$ (Eq.\ \ref{eq:time_to_peak}) as a function of time, not number of detections.

The median difference from the final estimate of the peak is shown as the solid black line in Fig.\ \ref{fig:sn_timing}.
The gray shaded region is contains the central $65\%$ of the difference measurements.

At time $t=t_{0}^{\ast}$, the median difference $t_{0}-t_{0}^{\ast}$ is less than one day.
We also note that at $t=t_{0}^{\ast}$, there is a spread in the difference values, $t_{0}-t_{0}^{\ast}$.
The central $68\%$ of these difference are within $-2.1$ to $+1.3\text{ days}$.
We interpret this as a representative uncertainty on the SALT $t_0$ estimate at $t=t_{0}^{\ast}$ (for any candidate with peak magnitude brighter than $19\text{ mag}$).
We expect that the asymmetry is due to the fact that the peak is more easily constrained when the lightcurve has begun to `turn over''.

\cite{Moller2021} show that on average, \textsc{fink}'s first classification of type Ia SNe is six days before peak brightness using ZTF data.
At $t-t_{0}^{\ast}=-6.0$, SALT fits underestimate $t_{0}$ by around two days on average.
That is, six days before the `true' peak, the best available SALT fit estimates that the peak will be in only four days.
However, we expect that the first correct classification from \textsc{fink} will occur at an earlier phase with LSST data, given LSST's increased depth.

\begin{figure}
    \centering
    \includegraphics[width=\columnwidth]{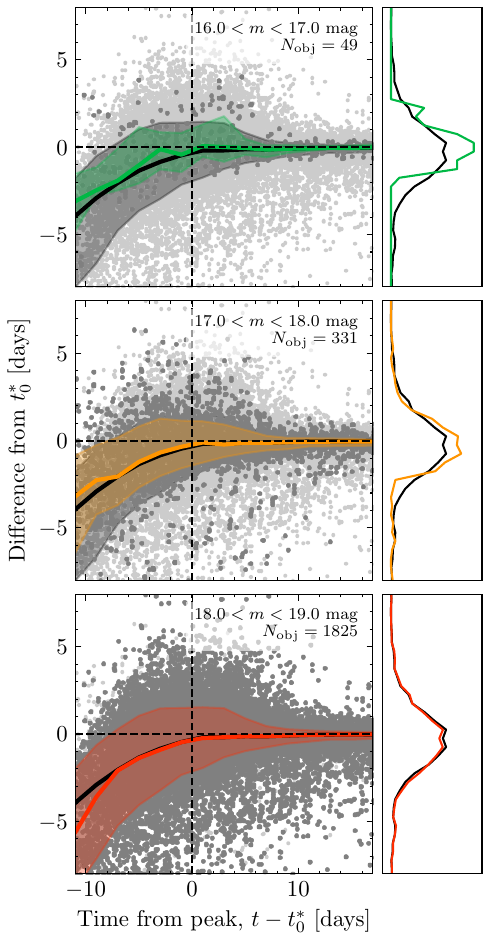}
    \caption{
        The convergence of the SALT zero-phase parameter $t_{0}$, as a function of SN phase, separating candidates by their maximum model $g$ measurement. 
        Large, dark grey points are the $t_{0}-t_{0}^{\ast}$ differences as a function of time from $t_{0}^{\ast}$ for candidates in that magnitude bin. 
        The coloured solid line and shaded region in each panel show the median and central $68\%$ region of the candidates in that magnitude bin. 
        The light grey points, median in black and $68\%$ region in grey for the entire sample are shown as in Fig.\ \ref{fig:sn_timing}.
        The auxiliary axes to the left of the main panels are the normalised distribution of differences $t_{0}-t_{0}^{\ast}$ in the $\text{MJD}-t_{0}^{\ast}$ interval $\left[-1,+1\right]$.
        The black curve (the same in each panel) is for the full sample.
    }        
    \label{fig:sn_timing_mag}
\end{figure}

Fig.\ \ref{fig:sn_timing_mag} shows the same statistic as Fig.\ \ref{fig:sn_timing}, but with candidates separated by their brightest detection in ZTF-$g$, in three magnitude bins.
The behaviour is similar for all three bins.
For the supernovae in the brightest magnitude bin, the $68\%$ confidence interval is narrower (from $-1.5$ to $+1\text{ days}$), but noisier around the time of the peak when compared with the fainter magnitude bins.
The the confidence interval and median difference as a function of time for the faintest subset (the lower panel of Fig.\ \ref{fig:sn_timing_mag}) is very similar to that of the whole subset, simply because there are far more candidates in this subset.

\subsubsection{Candidate rates}

\begin{figure}
    \centering
    \includegraphics[width=\columnwidth]{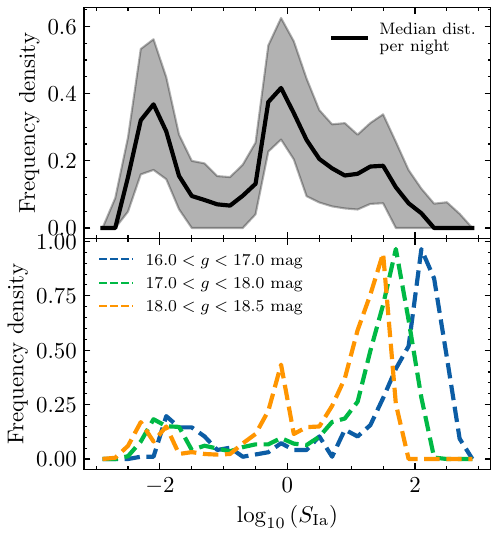}
    \caption{
        Upper panel: The median (black line) of the normalised distributions of all SNe Ia scores per night, $\log_{10} \left(S_{\rm Ia}\right)$.
        The shaded grey region shows the central $68\%$ of the distributions.
        Lower panel: the normalised distribution of scores of candidates within one day of $t_{0}^{\ast}$ (that is, candidates $|t - t_{0}^{\ast}| < 1$).
    }
    \label{fig:log_score_dist}
\end{figure}

We simulate how AAS2RTO would perform on the ZTF archival data we describe above, by processing alerts in chronological order and calculating the score which would have been available on that day (24 hour period).
In this test, we use the full set of alerts which we have scraped from \textsc{fink} instead of the smaller, good quality sample of 2205 SNe lightcurves, as this is how AAS2RTO will operate in real-time.
That is, we will not know in advance (at a point part way through the evolution of any given SN candidate) whether the final lightcurve will meet the quality cuts.

For each night with a non-zero number of alerts, we find the distribution of the log of scores $S_{\rm Ia}$ (the SNe Ia score without considering candidate visibility; Eq.\ \ref{eq:score_sneia}).
The median of these distributions is shown in Fig.\ \ref{fig:log_score_dist}, along with the central $68\%$ range.
The shape of this median distribution is mainly influenced by the factors $x_{\rm mag}$ and $x_{\rm peak}$.
The peak at $\log_{10}S_{\rm Ia}\sim-2$ is due to the minimum allowed value of $x_{\rm peak}=10^{-2}$, when candidates are very far from the apparent peak.
On average, a large fraction of candidates have a score below the base score of $S_{\rm base}=1$ (that is, they have $\log_{10}\left(S_{\rm Ia}\right)<0$).
These are candidates which will be penalised due to being faint, far from the estimated model peak, declining, or old.
The candidates with $\log_{10}S_{\rm Ia}>0.0$ which will have been promoted for the opposite reasons.

Fig.\ \ref{fig:log_score_dist} also shows distributions of $\log_{10}\left(S_{\rm Ia}\right)$ for any candidate within one day of the final peak estimate $t_{0}^{\ast}$, separated by magnitude (lower panel).
These distributions contain all $S_{\rm Ia}$ values from the duration of the simulated flow of alerts, as we are unable to present the median distribution per night due to the small number of candidates on average per night which meet these criteria (see Fig.\ \ref{fig:candidate_rates} below).
On the whole, these distributions of $S_{\rm Ia}$ for the candidates we aim to select behave as designed.
The candidates in the brightest subset within one day of the peak have higher values of $S_{\rm Ia}$ on average.
There are a few candidates within one day of $t_{0}^{\ast}$ which have very low values of $S_{\rm Ia}$, likely where the best SALT fit available at $t\sim t_{0}^{\ast}$ is poor.
Although Fig.\ \ref{fig:sn_timing_mag} shows that on average the best available estimate of $t_{0}$ is accurate at $t=t_{0}^{\ast}$, there is still some scatter.
We expect that the smaller peak at $\log_{10}\left(S_{\rm Ia}\right)\sim0$ for the faintest subset is due to candidates which have insufficient detections before the peak for a SALT fit at $\text{MJD}=t_{0}^{\ast}$, and so are assigned the default value of $x_{\rm peak}=1.0$.

On days with a non-zero number of alerts, we count the number of supernovae within half, one and two days of the best available value of $t_{0}$, and plot the frequency distribution in Fig.\ \ref{fig:candidate_rates}, for candidates which have magnitude $m<18.5\text{ mag}$.
That is, we count the number of candidates with $|t - t_{0}|< 1\text{ day}$.
We only include days when ZTF alerts were broadcast (for example, we do not include the large gap between December 2021 and February 2022 when there were no ZTF alerts broadcast by \textsc{fink}).

\begin{figure}
    \centering
    \includegraphics[width=\columnwidth]{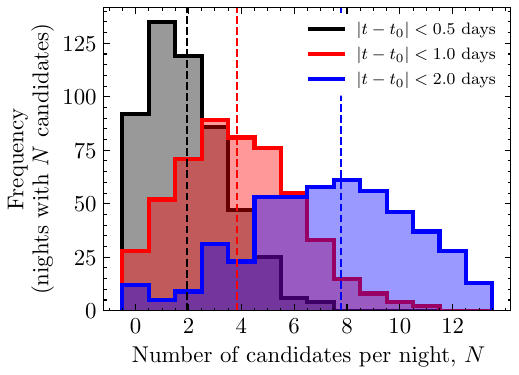}
    \caption{
        Frequency of number of candidates per day ($24\text{ hour}$ period) within a half, one and two days of the current predicted peak (in black, red and blue respectively).
        These measurements are made only for days where ZTF alerts were delivered.
        The mean number of candidates per day for each selection is denoted by the dashed vertical line.
        There are a mean of $\bar{N}=2.08\pm0.07$ candidates within 0.5 days of the peak, $\bar{N}=4.2\pm 0.1$ per night within one day of the predicted peak, and $\bar{N}=8.3\pm0.2$ within two days of the predicted peak. }
    \label{fig:candidate_rates}
\end{figure}

From Fig.\ \ref{fig:candidate_rates}, we expect an average of $\bar{N}=4.4\pm0.1$ SNe which are within one day of peak brightness, and $\bar{N}=8.7\pm0.1$ SNe within two days of peak brightness.
We choose to report the number of candidates $N$ per day using the current value of $t_{0}$ (not the final value, $t_{0}^{\ast}$), as this is the information that AAS2RTO will have when it operates in real-time.
For this same reason, we do not exclude candidates which have later been spectroscopically classified as non-Ia supernovae.

We use the values of $\bar{N}$ only to demonstrate the number of potential candidates for an observing campaign aided by AAS2RTO, and do not intend that they should be read as a measurement of the cosmological rate of Type Ia supernovae.
However, it is still useful to compare the mean number of SNe Ia candidates we could expect from ZTF to the number that could be expected based on estimates of SNe Ia rates.
To compute a simple estimate of the expected number of supernovae in the ZTF footprint, we assume a characteristic absolute magnitude of an SN Ia as $\mathcal{M}_{B}=-19.3\text{ mag}$, motivated by the peak of the SNe Ia luminosity function \citep[e.g.,][]{Perley2020, Desai2024}.
For simplicity, we assume the same $B$-band apparent magnitude as the $i$-band limit, $B_{\rm lim}=18.5$.
Combining these two values results in a distance modulus limit of $\mu=37.8\text{ mag}$, corresponding to $z=0.08$.
Using Palomar Transient Factory data, \cite{Frohmaier2019} compute the volumetric rate of SNe Ia ($r_{\rm V}$, the number of SNe Ia per co-moving volume, per year), and obtain $r_{\rm V}=2.43\pm 0.29\times10^{-5}\text{ SNe}\text{ Mpc}^{-3}\text{ yr}^{-1}\text{ h}_{70}^{3}$ (statistical uncertainty) at redshift $z<0.08$.
\cite{Perley2020} estimate that the effective area of the ZTF public survey which has been observed with at least three day cadence is $14,400\text{ deg}^{2}$.
From the rate $r_{\rm V}$ quoted above, we should expect $\sim1750$ supernovae per year in this ZTF footprint at $z<0.08$, or approximately $3.72\pm0.44$ per 24 hours.

The value which we can most fairly compare with $r_{\rm V}$ is the number of candidates within $\pm0.5$ days of the peak (an interval of 24 hours), $\bar{N}=2.08\pm0.07$.
This is a factor of two smaller than the expected $3.72\pm0.44$ supernovae per 24 hours estimated with the measured SNe Ia rate, $r_{\rm V}$.
We attribute this to missed SNe, and poorly-sampled lightcurves which do not meet the quality cuts when computing the score $S_{\rm Ia}$, and so are excluded from our ranked lists.
We expect that the completeness of the \textsc{fink} classifier only plays a small role.
\cite{Moller2020} show the \textsc{fink} Ia-vs-non-Ia classifier is capable of correctly classifying $86\%$ of SNe at two days before the SN Ia peak.

\begin{table}[]
    \centering
    \caption{
        Fraction of candidates from the good quality sample of 2205 ZTF lightcurves which appear in the top 3 and top 5 of ranked lists from AAS2RTO. 
        $^{a}$The percentage of candidates from the good quality sample of 2205 ZTF lightcurves which appear at least once in the top $N$ candidates of a ranked list.}
    \begin{tabular}{lccc}
        \hline
        Magnitude range & \# of SNe & \% in top 3$^{a}$ & \% in top 5$^{a}$\\
        \hline
        $16<g<17$ & 49 & 77\% & 82\% \\
        $17<g<18$ & 331 & 37\% & 53\% \\
        $18<g<18.5$ & 560 & 7\% & 19\% \\
        \hline        
    \end{tabular}
    \label{tab:sn_stats}
\end{table}

Finally, we consider the number of candidates which are ranked highly and are also in our sample of good quality SNe lightcurves (summarised in Table \ref{tab:sn_stats}).
We count the number of SNe from this sample which occur at least once in the top three or top five ranked candidates output from AAS2RTO.
For instance, 38 out of the 49 in the brightest subset of good quality SNe lightcurves ($16<g<17\text{ mag}$) appear in the top five ranked candidates at least once, or $81\%$.
This compares with $53\%$ in the second brightest subset, and only $19\%$ in the faintest subset.
We expect that this is primarily driven by the factor $x_{\rm mag}$, which promotes brighter candidates.
However, we also note that the fraction of lightcurves with peak $16<g<17\text{ mag}$ that are included in the good quality ZTF sample is larger than the fraction of lightcurves $17\text{ mag}<g<18\text{ mag}$ included, and similarly for the $18\text{ mag}<g<18.5\text{ mag}$ subset.
That is, there is a larger fraction of faint candidates which are included in ranked lists, but do not meet the stricter criteria we use for evaluating the use of SALT lightcurves to estimate the time of peak brightness.
We note that here the faintest subset is shallower than in Fig.\ \ref{fig:sn_timing_mag} ($18<g<18.5\text{ mag}$ instead of $18<g<19\text{ mag}$), as candidates with $g>18.5\text{ mag}$ are automatically excluded from ranked lists.

\subsection{Simulating LSST SNe Ia observations}

To investigate how useful our scoring function will be for prioritising SNe Ia candidates discovered by LSST, we use the LSST Operations Simulator tool\footnote{OpSim: \url{https://github.com/lsst/rubin_sim}} \citep[OpSim,][]{Delgado2014} to build a simple simulation of SNe Ia observations.
We use OpSim's \texttt{baseline} simulation, a realistic ten-year simulation of LSST visits providing an observation schedule with celestial coordinates, timestamp, filter, and $5\sigma$ depth for each visit.
The \texttt{baseline} simulation also includes a realistic number of nights where no observations are taken, based on the expected observing conditions at Cerro Pach{\'o}n. 

In 100 redshift shells spaced logarithmically from $0<z<0.4$, we compute the expected number of SNe Ia per volume using the volumetric rate $r_{V}$ from \cite{Dilday2008},
\begin{equation}
    r_{V} = 2.6\times10^{5}\left(1+z\right)^{1.5} \text{SNe}\text{ Mpc}^{-3}\text{ yr}^{-1},
\end{equation}
using the midpoint redshift $z$ of the shell.
For each shell, we draw the number of simulated SNe from a Poisson distribution with a mean $r_{V}\left(z\right) \times V\left(z\right) \times L$, where $V\left(z\right)$ is the comoving volume of the redshift shell and $L$ is the length of the survey measured in years.
We choose $L=5\text{ yr}+30\text{ days}$.
We assign a random zero-phase parameter $t_{0}$ to each supernovae, distributed uniformly from 30 days before the first simulated LSST visit, to the length of the survey $L$.
The extra 30 days ensures that there are SNe which are SNe which are declining in brightness at the time of the first simulated visit.
Simulated SNe in each redshift shell are assigned a redshift $z$ distributed uniformly across the shell, and a random pair of coordinates distributed uniformly on the celestial sphere.
We assign SALT parameters $x_{1}$ and $c$ following the distributions from \cite{Taylor2021} using the DES survey (see Fig.\ \ref{fig:salt_dist}).

To determine the SALT amplitude parameter $x_{0}$, we consider the Tripp relation \citep{Tripp1998},
\begin{equation}
    \label{eq:tripp}
    \mu\left(z\right)=m_{B} - \mathcal{M}_{B} + \alpha x_{1} - \beta c,
\end{equation}
where $\mu\left(z\right)$ is distance modulus, $\mathcal{M}_{B}$ is the absolute $B$-band magnitude of SNe Ia, $m_{B}$ is the apparent $B$-band magnitude, and $\alpha$ and $\beta$ are the parameters which represent the slopes of the stretch-luminosity and colour-luminosity relations.
We use $\alpha=0.148$ and $\beta=3.1$ as reported in \cite{Brout2022} using the Pantheon+ sample (their Table 2).
Simulated SNe Ia are assigned an absolute magnitude $\mathcal{M}_{B}$ using a normal distribution centred at mean $\mathcal{\bar{M}}_{B}=-19.3\text{ mag}$, with width $\sigma=0.15\text{ mag}$.
Using the Planck cosmological parameters \citep{Planck2020} to compute $\mu\left(z\right)$, we rearrange Eq.\ \ref{eq:tripp} to find the apparent magnitude $m_{B}$.
For the SALT templates in the \textsc{sncosmo} framework, the amplitude $x_{0}$ is related to apparent peak magnitude as $m_{B}=-2.5\log_{10}\left(x_{0}\right)+10.5$.

For each simulated LSST visit provided by OpSim, we `measure' the flux of simulated SNe Ia (using \textsc{sncosmo}'s implementation of the SALT models) at the observation timestamp in the appropriate filter, for each SN Ia in the field of view of $9.4\text{ deg}^{2}$.
We calculate the true apparent magnitude from the SALT model flux, and use the S/N estimate to obtain an uncertainty estimate (on both flux and magnitude).
We then obtain `observed' apparent magnitudes by sampling a single value from a Gaussian centred at the true apparent magnitude, with a width equal to the magnitude uncertainty estimate.

We note that in order to measure fluxes for all six LSST $ugrizy$ bands, we must use SALT3 models \citep{Kenworthy2021} instead of SALT2.
This is because SALT2 is trained on data $2000$--$9200\,\AA$, and is therefore unable to measure the flux in the LSST $z$- and $y$-bands ($8030$--$9385\,\AA$ and $9084$--$10945\,\AA$ respectively).
However, \cite{Kenworthy2021} show (using a Kolmogorov-Smirnov test) that the joint distributions of $x_{1}$ and $c$ for SALT2 and SALT3 models are statistically indistinguishable.

\begin{figure}
    \centering
    \includegraphics[width=\columnwidth]{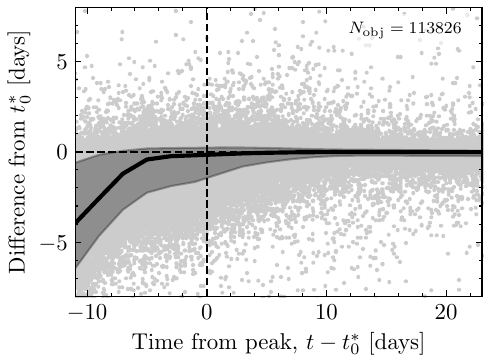}
    \caption{
        Convergence of estimate of time to peak, $t_{0}-t_{0}^{\ast}$ for simulated LSST SNe Ia lightcurves.
        Here, $t_{0}^{\ast}$ is the zero-phase parameter used as input to generate the simulated observations.
        We have plotted only $5\%$ of the points in light grey.
    }
    \label{fig:sn_timing_lsst_baseline}
\end{figure}

In Fig.\ \ref{fig:sn_timing_lsst_baseline}, we show the time to peak convergence plot using these simulated lightcurves (shown for ZTF in Fig.\ \ref{fig:sn_timing}).
We refit the first $N$ simulated detections in the lightcurve, and compute the difference between the latest estimate of the zero-phase parameter, $t_{0}$, and the true value $t_{0}^{\ast}$ (which was input to the simulated SNe).
We apply the same quality cuts as for the ZTF data (described in Sect.\ \ref{sec:ztf_archive}).
However, given LSST's significantly increased depth compared with ZTF, we use a limiting magnitude of $m_{B}<21.0\text{ mag}$.
Fig.\ \ref{fig:sn_timing_lsst_baseline} shows that the estimates of $t_{0}$ with increasing phase converge to the true value earlier than that of the real ZTF data, with a smaller spread.

\begin{figure}
    \centering
    \includegraphics[width=\columnwidth]{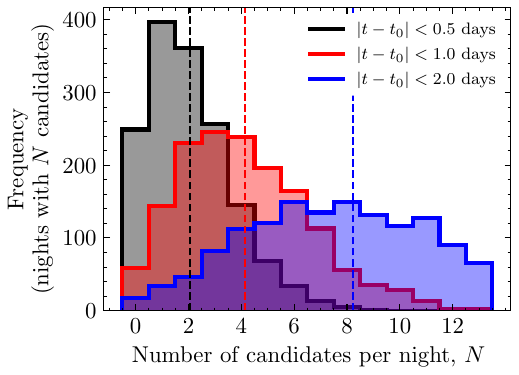}
    \caption{
        The number of expected candidates within half, one and two days of the peak from simulated LSST observations.
        There are a mean of $2.05\pm0.04$ within 0.5 days of the true peak, $4.12\pm0.06$ within one day of the peak, and $8.2\pm0.1$ within two days of the peak (indicated by dashed vertical lines).
    }
    \label{fig:candidate_rates_lsst}
\end{figure}

Finally, Fig.\ \ref{fig:candidate_rates_lsst} shows the number of simulated LSST candidates per night which are within half, one and two days of the true (simulation input) value of $t_{0}$ with a peak $g<18.5\text{ mag}$.
We note here that we use the number of candidates within (e.g.) one day of the \textit{true} value of the zero-phase parameter, $t_{0}^{\ast}$.
The main reason for this is that the accuracy of the time to peak estimates presented in Fig.\ \ref{fig:sn_timing_lsst_baseline} will likely be different when using `real' LSST data (see below for more discussion of assumptions made in this section).
That is, although the number of input SNe is realistic, the final LSST observation cadence is not fixed, and will therefore have some impact on Fig.\ \ref{fig:sn_timing_lsst_baseline}.
As with the similar statistic from ZTF, we only compute the number of nights in the simulation where OpSim provides simulated observations.
We find that we select $2.05\pm0.04$ within 0.5 days of the true peak, $4.12\pm0.06$ within one day of the true peak, and $8.2\pm0.1$ within two days of the true peak.

We note the results in this section are based on an very simple and idealised `toy model' of LSST supernovae observations, with the impact of our assumptions described here.
First, we do not include other types of SNe in our simulation, whose lightcurves are poorly explained by SALT models.
Therefore, there is no `contamination' by non-Ia SNe types.

We compute the uncertainties on measured fluxes and magnitudes using OpSim's point source signal-to-noise ratio (S/N) estimation tool.
This tools estimates the S/N using only the observed magnitude of a point source and $5\sigma$ depth.
However, this S/N is for a single-epoch science image.
In reality, SNe fluxes are measured with from difference images (the single-epoch image minus a deep coadd image), so that the flux contribution from the host galaxy has been removed.
As we do not consider host galaxies in our simulated SNe Ia sample, we do not make this subtraction step, so our SNe flux uncertainties will be underestimates.
Additionally, this means that we do not account for the possibility that the earliest stages of an SN Ia lightcurve are masked by their host galaxy (i.e., are non-detections).

When simulating the alerts from LSST, we assume that every photometric detection is a `valid' alert, and would be correctly classified as a supernovae.
These assumption will have very little impact on our results, because the broker classifiers have a high classification accuracy by the time SNe Ia are at peak brightness.
For a score designed to promote very young supernovae, this assumption may have more of an impact, as the LSST broker classifiers do not have as high a recall (true positive rate) at early lightcurve phase.

\section{AAS2RTO's place amongst schedulers}\label{sec:comparison}

\subsection{Greedy strategy}

The concept of greedy algorithms for telescope scheduling that we have implemented for AAS2RTO is not a new one. 
An advantage of greedy schedulers is that they are extremely flexible, and can respond rapidly to new data or conditions.
The original scheduler at the 2m robotic Liverpool Telescope at La Palma is described as a `dispatch' scheduler \citep{SteeleCarter1997}.
It uses an `efficiency' function (written in Perl), which considers altitude, site conditions and a prescribed values of scientific priority, which our factors and scoring function is similar to.
The scientific priority values are provided for each observation target when it is submitted by Telescope Allocation Groups.
The scientific priorities are normalised so that the overall priority  of the set of observations for each Telescope Allocation Group is equal.
AAS2RTO differs in that a candidate's priority value (score) can vary with time, as fixed priority values are less useful for transient objects which evolve on short timescales.
Further, as AAS2RTO is primarily designed to prioritise targets according to a single scientific goal (that is, not for a facility with competing observing programmes from many groups), we do not carry out the priority normalisation step of \cite{SteeleCarter1997}.

More recently, ``the Conductor'' scheduler \citep{Andersen2019} for the Stellar Observations Network Group\footnote{also `Stellar Oscillations Network Group'} \citep[SONG,][]{Grundahl2008} first makes a broad prioritisation of their `targets' with a pre-determined list of categories.
These categories have a priority which decreases with how critical the timing of the next observation is.
The first category contains targets which \textit{must} be observed at a specific time, followed by targets which require many observations at at regular intervals, and then targets which require few observations which can be taken at any time.
Targets in each of these categories are then ranked according to priorities calculated similarly to our own factors.
They consider the required time interval between observations and the time since the last observation, and pre-assigned scientific priorities by a Time Allocation Committee.
Targets in the first category are observed until this list is exhausted, then observing continues with the highest ranked in the second category.
As AAS2RTO is designed for prioritisation of candidates based on alerts distributed in real-time, we have decided against using a semi-automated pre-categorisation strategy like SONG.

AAS2RTO is also similar in approach to the scheduler for the Gravitational-wave Optical Transient Observer \citep[GOTO,][]{Dyer2018, Dyer2020}.
The aim of GOTO is to cover the large `probability regions' for finding optical counterparts of gravitational wave signals and Gamma Ray Bursts (GRBs) with imaging `tiles' (telescope pointings).
This aim is very different to the example AAS2RTO case we present in Sect.\ \ref{sec:test_case}.
Nevertheless, there are similarities in prioritisation strategy.
The GOTO scheduler begins with a predefined `starting rank' for a candidate, which our `base score', $S_{\rm base}$, can be compared with.
The starting rank for each tile covering a probability regions is based on the astrophysical type (gravitational wave event from black hole and/or neutron star merger, or a GRB), and detection instrument.
The rank is then modified by considering the time since a tile was last observed, the fraction of probability it contains, and the airmass of the tile at its current altitude.
Each of these is considered independently, as with our factors, $x_{i}$.

The primary purpose of the SOXS instrument is to conduct spectroscopic follow-up observations of astrophysical transients in the UV-visible and near-infrared.
The SOXS scheduler scrapes data from ZTF, ATLAS and other surveys in a similar manner to AAS2RTO to ingest new candidates and update the existing candidate pool.
In contrast to AAS2RTO, the SOXS scheduler \citep{Asquini2022, Asquini2024} is not designed to schedule observations for a single science case, meaning that assigning scientific priorities to candidates is not as straightforward.
These priorities are therefore assigned and updated by a SOXS scientific committee on a daily basis.
The SOXS scheduler then takes the assigned priorities, and proposes an observing schedule which maximises the number of candidates observed with high scientific priority, while also respecting any observing constraints (visibility, moon separation).
The scientific committee still has a `veto' power on the proposed schedule, and modifies it to account for any additional targets of opportunity.
We expect that for the first few months of operation of LSST (at least), we will use the output ranked lists from AAS2RTO in a similar manner to SOXS, using them initially as a `proposed schedule'.
This will allow us to make any necessary adjustments to the scoring function, or data ingestion steps (e.g. due to any unforeseen differences between ZTF and LSST alerts), before the DK1.54m is operated in a more and more autonomous manner.

\cite{Rehemtulla2024} introduce \texttt{BTSbot}, a new tool which automates identifying new bright candidates for the ZTF BTS spectroscopic follow-up survey.
\texttt{BTSbot} uses a convolutional neural network trained on the images from ZTF alerts, along with features extracted from ZTF lightcurves to provide a `bright transient score' for individual ZTF alerts.
\texttt{BTSbot} performs with comparable purity to human `scanners' who manually select and trigger spectroscopic observations.
At present, AAS2RTO does not use any deep learning modules (although many alert streams from brokers are filtered using deep learning).
However, it would be useful to produce individual factors using deep learning.
For instance, a factor could be designed around a deep learning model which predicts the time of a candidate SN Ia peak brightness, and reserve fitting expensive SALT model fitting for the most promising candidates.

Very large projects with a variety of scientific goals often prefer not to use a greedy strategy as they do not provide a `globally' optimum schedule.
For instance, ZTF \citep{Bellm2019a} describe their aim of maximising transient discovery rate, and the need to balance the total number of exposures taken, the cosmological volume probed, and cadence.
Similarly, LSST's Deep-Fast-Wide survey has many scientific aims, and will need to balance transient discovery as well as overall depth and homogeneity over the whole survey footprint.
As a counterexample, however, the Science Planning and Scheduling System for the \textit{Hubble Space Telescope} used a greedy search algorithm successfully during the first seven years of operation \citep{Samson1998}.

The Target and Observation Manager Toolkit \citep[TOMToolkit,][]{Street2018} has a different aim to the tools mentioned above in that it is a general-purpose toolkit.
It can be used to build whole observation management systems from scratch, or used as a library of useful tools.
We aim in future to use some of the functionality of TOMToolkit to improve AAS2RTO.
In particular, we will use the alert and data management modules.
Currently, AAS2RTO stores photometric data from ZTF/LSST brokers and ATLAS and transient listings from TNS very simply using \textsc{ascii} files.
Although this is not the most efficient approach, it is easiest to implement.
This works well for the narrow science case we have described in Sect.\ \ref{sec:test_case}, but could become unwieldy for broader science cases where there are more candidates.
Further, AAS2RTO is currently controlled only with command-line interface.
The TOMToolkit allows for Graphical User Interfaces (GUIs) to be easily implemented, which would make AAS2RTO more user-friendly.

    Finally, we compare AAS2RTO with schedulers which use a (mixed) integer linear programming solution (MILP/ILP).
    MILP and ILP solutions have been shown to provide globally optimal solutions for extremely complex scheduling problems.
    For instance, \citet{Solar2016} demonstrate how a MILP solution can provide an optimal observing schedule for an entire six-month ALMA observing season, while still maintaining the flexibility to adapt in the short-term.
    Like other schedulers for observatories with many competing observing programmes (often containing several targets), fixed scientific priorities for each programme are assigned by science committees.
    \citeauthor{Solar2016}'s ALMA scheduler uses a commercial MILP solver (IBM-ILOG-CPLEX) to maximise the total `scientific throughput' of the observatory (the sum of the scientific priorities of programmes observed) in the long-term, while also minimising telescope idle time in the short-term and preferring completed programmes.
    Similarly, \citet{Lampoudi2015} use an ILP solution to schedule the Las Cumbres Observatory Global Telescope (LCOGT) network, with a very large number of telescopes of different sizes and capabilities.
    Requests for observations with LCOGT are made on a twice-yearly basis, with the additional complication that requests can ask for any subset of the available telescope resources.
    \citeauthor{Lampoudi2015}'s scheduler maximises the sum of scientific priorities of observed targets, and is shown to provide a close-to-optimal solution when the network is over-subscribed (that is, more observing time is requested than is available).
    The LCOGT scheduler uses the commercial Gurobi ILP solver.
    The ZTF scheduler described above \citep{Bellm2019a} also uses the Gurobi ILP solver.
    As AAS2RTO is designed for observing candidates which are known only within days of observation (SNe Ia with a rise time of $\sim19$ days), and not for competing proposals which are submitted before an observing season, the simple greedy scheduler allows for sufficient flexibility in the short-term.

\subsection{Candidate visibility considerations}

One criticism of greedy algorithms in the telescope schedulers is that they do not sufficiently account for the visibility of candidates, and risk missing maximal scientific output by missing candidates which will quickly set below the horizon.
\cite{Rana2017} raise this issue in the context of covering the large ($\gtrsim100\text{ deg}^{2}$) probability regions for gravitational wave event follow-up. 
They demonstrate that greedy schedulers can lead to less than optimal coverage compared with their best-performing algorithm due to parts of the probability region setting before observation.
\cite{Rana2017} point out, however, that complicated observing schedules can give \textit{worse} performance than greedy algorithms in case of unreliable instrumentation or weather.
\cite{SteeleCarter1997} state this last argument as their reason for implementing a `dispatch' scheduler for the Liverpool Telescope at La Palma, where the observing conditions can change rapidly and frequently.

We partially address concerns about target visibility using the visibility factor in Eq.\ \ref{eq:x_vis} (Sect.\ \ref{sec:candidate_visibility}).
    Our approach is similar to that taken by the ACP scheduler \citep[ACPS][]{Denny2004, Denny2006}.
    ACPS uses an efficiency function modified from \citet{SteeleCarter1997}, with an additional term which accounts for the visibility of targets.
    This `highest altitude' term is designed to prefer targets at their transit altitude, and accounts for the fact that some rising targets will never reach transit altitude.
    Additionally, ACPS allows for repeat observations of the same target at a specified time interval.
    ACPS therefore considers the centroid altitude of the repeat observations at their specified time interval.
    Given the motivation of rapid response spectroscopic follow-up of primarily newly discovered transients or specifically time-targeted observations with AAS2RTO (which are the main focus in this work), we do not include automated repeating observations of the same target at this point, but can be considered in future upgrades of AAS2RTO.
Further, our visibility factor is a similar approach to the SONG's Conductor scheduler, which gives preference to setting targets over rising targets by using fixed weights.

In future developments to AAS2RTO, we hope to address this issue further by extending the concept of the score to the entire night.
Considering combinations of observations throughout the night would allow the for a selection of observations which maximise the score for the entire night, but would result in a more rigid schedule.
This option would enable additional use cases for AAS2RTO, but would not provide the flexibility of the existing greedy implementation, which is useful for real-time rapid response observations. 

\section{Conclusions}

We discussed the implementation and greedy strategy of our new candidate prioritisation tool, AAS2RTO.
It will be used to prioritise transient candidates from LSST, in order to use the DK1.54m telescope at La Silla, Chile, as a spectroscopic follow-up facility for transient alerts from LSST. 

The greedy strategy used by AAS2RTO is described generically, along with the individual factors which constitute the final score.
The factors for each candidate can be weighted to preferentially promote candidates with specific properties.
As an example science case of AAS2RTO, we describe in detail the factors we use to prioritise candidate SNe Ia which are at peak brightness.
In short, these factors promote candidates which are bright, not much older than the typical SNe Ia rise time, and not declining in brightness.
Finally, the factor considers the remaining visibility of the candidate during an observing night, and prioritises candidates which will soon set below a permitted minimum altitude.

Using archival ZTF data, we simulate how AAS2RTO could perform in real-time.
We show that using SALT models, we are able to estimate the average epoch when peak brightness occurs with a precision of $\pm_{2.1}^{1.3}\text{ days}$.
We estimate that we would have around $\bar{N}=2.08\pm0.07$ SNe per night within 0.5 days of the peak, for a facility with the same magnitude limitations as the DK1.54m ($r>18.5\text{ mag}$).
This is around a factor of two lower than the actual number of supernovae we expect in the ZTF footprint, and we attribute this difference to the quality cuts we require for candidates.
We perform a similar experiment with simulated LSST alerts, and find that we can expect $2.05\pm0.04$ candidates from LSST within 0.5 days of the true peak.
Finally, we compare the greedy strategy of AAS2RTO with a selection of other available tools for observing.
Although many existing tools also use a greedy strategy for scheduling observations, none match exactly the example science case we have outlined here, or the flexibility of AAS2RTO.

\section*{Software}

The AAS2RTO project repository can be found at \url{github.com/aidansedgewick/dk154-targets-py38}.
Configuration instructions and technical details of the format of the user-supplied components (scoring functions and model building functions) are available in the \textsc{readme} section of the project repository.

In addition to tools described in the main text, this work made use of \textsc{python} \citep{VanRossum2009}, and the follow \textsc{python} packages:
    \textsc{astroplan} \citep{Morris2018},
    \textsc{astropy} \citep{AstropyCollaboration2022},
    \textsc{dustmaps} \citep{Green2018},
    \textsc{matplotlib} \citep{Hunter2007},
    \textsc{numpy} \citep{Harris2020},
    \textsc{pandas} \citep{McKinney2010},
    \textsc{scipy} \citep{Virtanen2020},
    \textsc{sncosmo} \citep{Barbary2016}.

\begin{acknowledgements}
    We are grateful to R.~A.~Street and M.~Hundertmark for helpful discussions.
    This work is supported by a Villum Young Investigator grant (project number 25501), a Villum Experiment grant (VIL69896) and research grants (VIL16599 and VIL54489) from VILLUM FONDEN.
    This work was made possible through the Preparing for Astrophysics with LSST Program, supported by the Heising-Simons Foundation and managed by Las Cumbres Observatory (proposal ID KSI-24, through grant 2021-2975).
    The recommissioning of DFOSC spectroscopic capabilities on the DK1.54m telescope has been supported by the Villum Experiment grant \textit{Cosmic Beacons} (project number 36225). 
\end{acknowledgements}

\bibliographystyle{aa} 
\bibliography{paper} 

\appendix

\section{Messaging examples}
\label{sec:messengers}

\begin{figure}
    \centering
    \includegraphics[width=0.8\columnwidth]{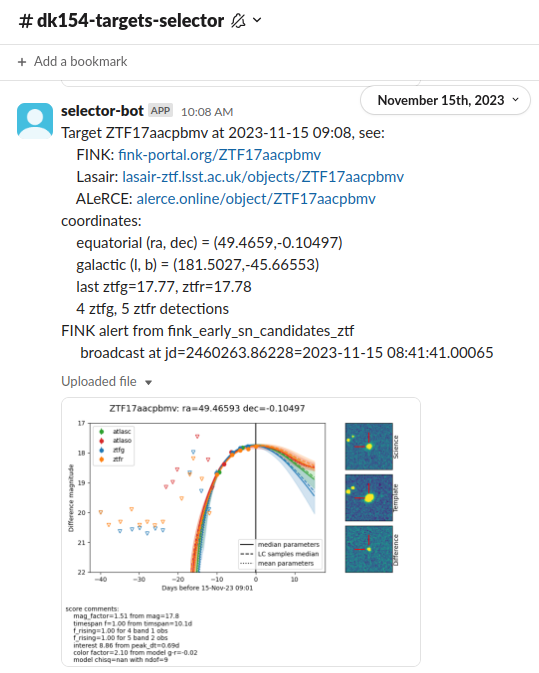}
    \caption{
        Messages sent using a Slack `bot'. 
        They contain lightcurve figures produced by AAS2RTO alongside some summary information about the candidate.
    }
    \label{fig:slack_message}
\end{figure}

\begin{figure}
    \centering
    \includegraphics[width=0.8\columnwidth]{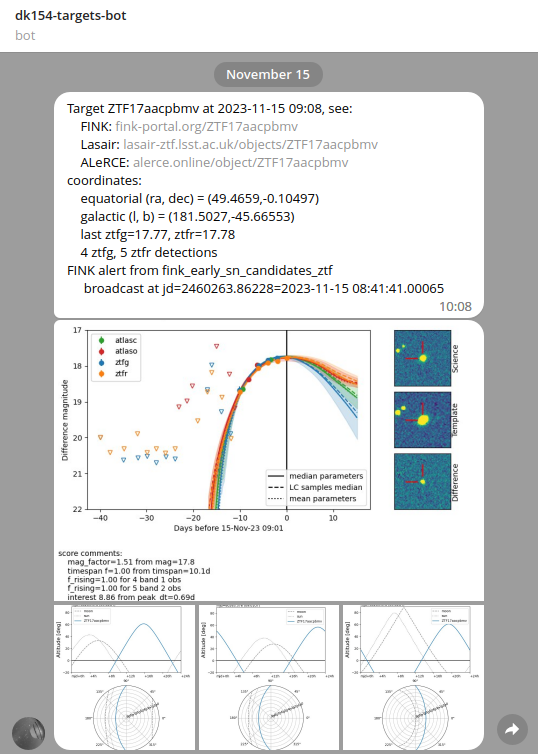}
    \caption{
        An example screenshot of messages sent using a Telegram `bot'. 
        In this example, altitude charts are also sent for a number of observatories in addition to the lightcurve and summary information about the candidate.
    }
    \label{fig:telegram_message}
\end{figure}

AAS2RTO is capable of sending messages about candidates to users through `bots'.
This is currently implemented with the team communication platform Slack (via the \texttt{slack\_sdk} API\footnote{Slack API: 
\url{https://slack.dev/python-slack-sdk/}}), and the instant messaging service Telegram (via the \texttt{python-telegram-bot} API\footnote{Telegram API: 
\url{https://python-telegram-bot.org/} (using synchronous version 13.12)}).
Fig.\ \ref{fig:slack_message} and Fig.\ \ref{fig:telegram_message} show examples of the Slack and Telegram messages sent from AAS2RTO, respectively.

The messaging functionality is supplementary to the main aim of providing a ranked list of candidates for observation.
This functionality is optional, as science cases or scoring functions which produce many candidates may produce an excessive number of messages.

\section{Behaviour of visibility factor}
\label{sec:vis_factor_appendix}

The visibility factor $x_{\rm vis}$ (defined in Eq.\ \ref{eq:x_vis}) does not have a maximum value due to the definition of $A_{\rm vis}$ (Eq.\ \ref{eq:A_vis}).
Rather, it is unbounded, with very large values when a candidates' altitude approaches $a_{\rm min}$.
This behaviour is undesirable if all the other factors for a candidate have been carefully constructed to be well-behaved in order to reflect the scientific aims of a user.
Here, we propose a slightly modified $\hat{x_{\rm vis}}$.

A simple solution is to define a maximum acceptable value, $A$, and choose the minimum value.
\begin{equation}
    \label{eq:xvis_hat_min}
    \hat{x_{\rm vis}} = \min\left(x_{\rm vis}, A\right).
\end{equation}

\begin{figure}
    \centering
    \includegraphics{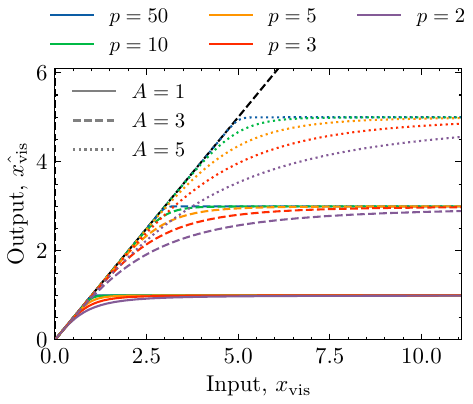}
    \caption{
        The saturation function of Eq.\ \ref{eq:xvis_hat_sat}, for combinations of the parameters $A$ and $p$, which set the maximum value, and the range of linear behaviour, respectively.
        The dashed black line is a one-to-one $\hat{x_{\rm vis}}=x_{\rm vis}$.
    }
    \label{fig:sat_func}
\end{figure}

An alternative proposal is to suppress the large values of $x_{\rm vis}$,
\begin{equation}
    \label{eq:xvis_hat_sat}
    \hat{x_{\rm vis}} = \frac{x_{\rm vis}}{\left[{1+\left(\frac{x_{\rm vis}}{A}\right)^{p}}\right]^{1/p}}.
\end{equation}
where $A$ is the maximum acceptable value of $x_{\rm vis}$, and $p$ is a constant $p>1$.
This is linear for small input values of $x_{\rm vis}$, but smoothly `saturates' to $A$ for large $x_{\rm vis}$.
The denominator of this `saturation' function is the $L^{\rm p}$-norm of the point (1, $x_{\rm vis}/A$).
For $x_{\rm vis}/A\ll 1$, $\hat{x_{\rm vis}}$ approaches zero, and the derivative is one (i.e., $x_{\rm vis}=\hat{x_{\rm vis}}$).
For $x_{\rm vis}/A\gg 1$, $\hat{x_{\rm vis}}$ approaches $A$.
For increasing $p$, the function output is approximately linear a larger ranges of input $x_{\rm vis}$.
As $p$ approaches infinity, the suggestion in Eq.\ \ref{eq:xvis_hat_min} is recovered.
The behaviour of Eq.\ \ref{eq:xvis_hat_sat} is shown in Fig.\ \ref{fig:sat_func}.

\begin{figure}
    \centering
    \includegraphics{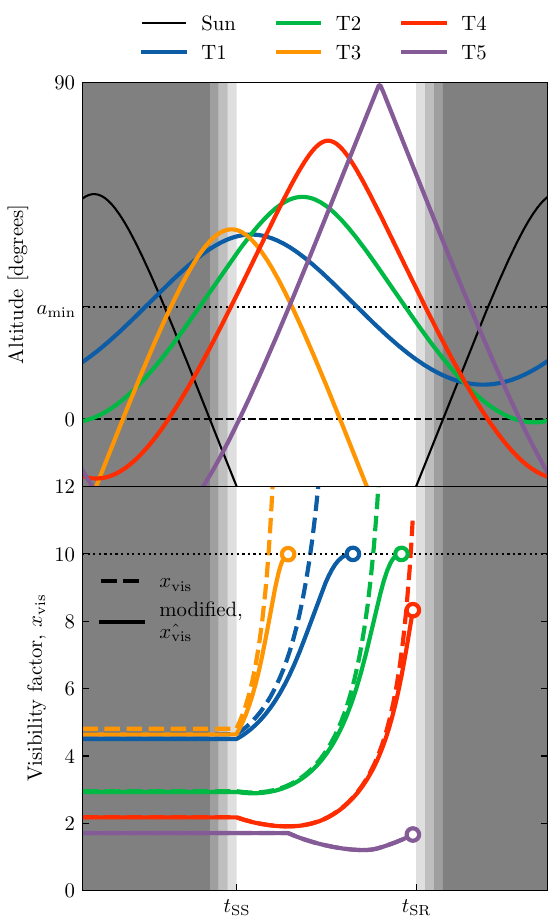}
    \caption{
        The visibility of several hypothetical candidates during a night is shown in the upper panel.
        The lower panel demonstrates how the visibility factor $x_{\rm vis}$ for each of these candidates changes as a function of time.
        The $x_{\rm vis}$ definition from the main body of the paper (dashed line) has no upper limit as candidates set below $a_{\rm min}$.
        The modified version (Eq.\ \ref{eq:xvis_hat_sat}, solid line) `saturates' to a maximum value.
        For this example, we have chosen $A=10$ and $p=10$, and $a_{\rm ref}=90^{\circ}$.
        The white region is when the Sun is below altitude $a<-18^{\circ}$ (astronomical night).
        Open circles mark when the time the candidate sets below the minimum acceptable altitude, $a_{\rm min}$, or the sun rises above $a=-18^{\circ}$ (whichever is earlier).
    }
    \label{fig:xvis_hat}
\end{figure}

The difference between the $x_{\rm vis}$ definition from the main body of the paper (Eq.\ \ref{eq:x_vis}), and the modified version in Eq.\ \ref{eq:xvis_hat_sat} is shown in Fig.\ \ref{fig:xvis_hat}, choosing parameters $A=10$ and $p=10$, $a_{\rm ref}=90^{\circ}$, and $a_{\rm min}=30^{\circ}$.
The visibility factor is shown for the same three hypothetical candidates in Fig.\ \ref{fig:altitude}, an additional candidate which sets close to sunrise ($t_{\rm SR}$) and a candidate which rises during the night and is still at high altitude at sunrise.
At small values of $x_{\rm vis}$, there is little difference between $x_{\rm vis}$ and $\hat{x_{\rm vis}}$.
For instance, for the candidate T5 in Fig.\ \ref{fig:xvis_hat}, the difference is not visible at any time.

We note that for observing time $t_{\rm obs}<t_{\rm SS}$ (during the day, grey shaded regions in Fig.\ \ref{fig:xvis_hat}), the visibility factor $x_{\rm vis}$ is constant due to the limits of integration in Eq.\ \ref{eq:A_vis}.

\end{document}